\documentclass[aoas]{imsart}

\RequirePackage{amsthm,amsmath,amsfonts,amssymb}
\RequirePackage[authoryear]{natbib}
\RequirePackage[colorlinks,citecolor=blue,urlcolor=blue]{hyperref}
\RequirePackage{graphicx}
\RequirePackage{gensymb} 
\RequirePackage{booktabs, array}
\RequirePackage{etoolbox}
\RequirePackage{pifont}
\RequirePackage[table]{xcolor}
\RequirePackage{colortbl}

\definecolor{bleudefrance}{rgb}{0.19, 0.55, 0.91}
\definecolor{brickred}{rgb}{0.8, 0.25, 0.33}
\newcommand{\cmark}{\textcolor{bleudefrance}{\ding{51}}}
\newcommand{\xmark}{\textcolor{brickred}{\ding{55}}}
\newcommand{\overbar}[1]{\mkern 1.5mu\overline{\mkern-1.5mu#1\mkern-1.5mu}\mkern 1.5mu}

\newcommand{\beginsupplementA}{%
	\setcounter{table}{0}
	\renewcommand{\thetable}{A\arabic{table}}%
	\setcounter{figure}{0}
	\renewcommand{\thefigure}{A\arabic{figure}}%
}

\newcommand{\beginsupplementB}{%
	\setcounter{table}{0}
	\renewcommand{\thetable}{B\arabic{table}}%
	\setcounter{figure}{0}
	\renewcommand{\thefigure}{B\arabic{figure}}%
}

\startlocaldefs

\endlocaldefs

\begin{document}

\begin{frontmatter}
\title{A Spatially Varying Hierarchical Random Effects Model for Longitudinal Macular Structural Data in Glaucoma Patients}
\runtitle{Spatially Varying Hierarchical Random Effects Model}

\begin{aug}
\author[A]{\fnms{Erica} \snm{Su}\ead[label=e1,mark]{ericasu@ucla.edu}},
\author[A]{\fnms{Robert E.} \snm{Weiss}\ead[label=e2,mark]{robweiss@ucla.edu}},
\author[B]{\fnms{Kouros} \snm{Nouri-Mahdavi}\ead[label=e3,mark]{nouri-mahdavi@jsei.ucla.edu}},
\and
\author[A]{\fnms{Andrew J.} \snm{Holbrook}\ead[label=e4,mark]{aholbroo@g.ucla.edu}}
\address[A]{Department of Biostatistics, Fielding School of Public Health, University of California Los Angeles, \printead{e1,e2,e4}}

\address[B]{Glaucoma Division, Stein Eye Institute, David Geffen School of Medicine, University of California Los Angeles, \printead{e3}}
\end{aug}

\begin{abstract}
	We model longitudinal macular thickness measurements to monitor the course of glaucoma and prevent vision loss due to disease progression. The macular thickness varies over a 6$\times$6 grid of locations on the retina with additional variability arising from the imaging process at each visit. Currently, ophthalmologists estimate slopes using repeated simple linear regression for each subject and location. To estimate slopes more precisely, we develop a novel Bayesian hierarchical model for multiple subjects with spatially varying population-level and subject-level coefficients, borrowing information over subjects and measurement locations. We augment the model with visit effects to account for observed spatially correlated visit-specific errors. We model spatially varying (a) intercepts, (b) slopes, and (c) log residual standard deviations (SD) with multivariate Gaussian process priors with Mat\'ern cross-covariance functions. Each marginal process assumes an exponential kernel with its own SD and spatial correlation matrix. We develop our models for and apply them to data from the Advanced Glaucoma Progression Study. We show that including visit effects in the model reduces error in predicting future thickness measurements and greatly improves model fit.
\end{abstract}


\end{frontmatter}



\section{Introduction}

Glaucoma damages the optic nerve and is the second leading cause of blindness worldwide \citep{kingman2004glaucoma}. As there is no cure, timely detection of disease progression is imperative to identify eyes at high risk of or demonstrating early progression so that timely treatment can be provided and further visual loss prevented. Ophthalmologists assess glaucomatous progression by monitoring functional changes in visual fields or structural changes in the retina over time. Visual field (VF) measurements assess functional changes by measuring how well eyes are able to detect light. Repeatedly measuring the thickness of retinal layers, such as macular ganglion cell complex (GCC), with optical coherence tomography (OCT) allows ophthalmologists to evaluate central retinal (macular) structural change over time. Both VF and OCT obtain data from multiple locations across the retina. In current practice, clinicians detect progression by modeling functional or structural changes over time using simple linear regression (SLR) for each subject-location combination \citep{gardiner2002examination, nouri2007comparison, tatham2017detecting, thompson2020comparing}. SLR does not accommodate the hierarchical structure that patients are members of a population and ignores the spatial arrangement of the data. For analyzing VF data at individual locations, \citet{montesano2021hierarchical} introduce a hierarchical model accounting for location and cluster levels fit to data from a single eye, \citet{betz2013spatial} and \citet{berchuck2019diagnosing} present models accounting for spatial correlation fit to data from a single eye, and \citet{bryan2017bayesian} describe a two-stage approach to fit a hierarchical model taking subject, eye, hemifield (one half of the VF), and location into account. While these methods exist for VF data, they cannot be directly applied to structural macular data as the measurement processes are markedly different. Key features of VF data that differ from structural data include censoring, heteroskedasticity, and a different underlying spatial structure.

We analyze data from the Advanced Glaucoma Progression Study (AGPS), a cohort of eyes with moderate to severe glaucoma. To monitor glaucoma progression, we model longitudinal macular GCC thickness measurements over a square $6 \times 6$ grid of 36 superpixels (roughly a $20\degree \times 20\degree$ area) for all subjects. For a single subject, the intercepts, slopes, and residual standard deviations (SD) vary spatially across superpixel locations. \citet{mohammadzadeh2021estimating} model GCC data from each superpixel separately and compare different Bayesian hierarchical models, preferring a model with random intercepts, random slopes, and random residual SDs. Our desired model needs to account for both the hierarchical structure of the data and the spatial correlations in both the population- and subject-level intercepts, slopes, and residual SDs and in the residuals. The parameters at the population level summarize information from the whole cohort at each superpixel location. Additional difficulties in modeling GCC data arise from the amount and sources of measurement error. Thickness measurements are reliant on automated segmentation algorithms, which may introduce spatially correlated errors unique to each imaging scan. We show that including visit effects to account for visit-specific errors reduces error in predicting future thickness measurements and greatly improves model fit. In this study, we motivate and develop the Spatially varying Hierarchical Random Effects with Visit Effects (SHREVE) model, a novel Bayesian hierarchical model with spatially varying population- and subject-level coefficients and SDs, accounting for spatial and within-subject correlation, between-subject variation, and spatially correlated visit-specific errors.

For the AGPS data, we allow the intercepts, slopes, and residual SDs to vary over space. Varying coefficient models are natural extensions to classical linear regression and extensively used in imaging studies and the analysis of spatial data \citep{hastie1993varying, ge2014analysis, zhu2014spatially, liu2019mixed}, where regression coefficients are allowed to vary smoothly as a function of one or more variables, and in our case, over spatial locations. Regression coefficients may vary over space in a discrete fashion as with areal units or in a continuous manner as with point-referenced data \citep{gelfand2010handbook}. In the context of imaging studies with grid data, a conditional autogressive (CAR) model  \citep{gossl2001bayesian, penny2005bayesian, ge2014analysis} or a Gaussian process (GP) model \citep{zhang2016bayesian, castruccio2018scalable} may be assumed for discrete or continuous spatial variation, respectively. In a GP model, coefficients from any finite set of locations has a multivariate normal distribution with a mean function and valid covariance function specifying the expected value at each location and covariance between coefficients at any two locations, respectively \citep{gelfand2010handbook}.

 \citet{gelfand2003spatial} first proposed the use of GPs to model spatially varying regression coefficients and multivariate Gaussian processes (MGP) for multiple spatially varying regression coefficients in a hierarchical Bayesian framework. We can assign GP priors at different levels in the hierarchy, which allows for flexible specification in hierarchical models \citep{gelfand2016spatial, kim2017hierarchical}. In our case with three components, spatially varying intercepts, slopes, and residual SDs, we employ MGPs to model the correlations between components within a location and across locations at both the subject and population level. MGPs are specified with a multivariate mean function and cross-covariance function, defining the covariance between any two coefficients at any two locations \citep{banerjee2003hierarchical}. For simplicity and computational convenience, separable cross-covariance functions are often used where components share the same spatial correlation and components within a location share a common covariance matrix, and the resulting covariance matrix is the Kronecker product of a covariance matrix between components and a spatial correlation matrix \citep{banerjee2003hierarchical}. Assuming all components share a common spatial correlation structure is likely inadequate in practice, as processes may be very different from each other in nature. Instead, we propose a nonseparable cross-covariance function to allow each process to have its own spatial correlation function.

Constructing valid cross-covariance models is a challenging task for nonseparable MGPs. \citet{genton2015cross} review approaches to construct valid cross-covariance functions for MGPs including the linear model of coregionalization \citep{wackernagel2013-bd, schmidt2003bayesian} and kernel and covariance convolution methods \citep{ver1998constructing, gaspari1999construction}. For univariate GPs, the Mat\'ern class of covariance models is widely used, featuring a smoothness parameter that defines the level of mean square differentiability and a lengthscale parameter that defines the rate of correlation decay \citep{guttorp2006}. \citet{gneiting2010matern} and \citet{apanasovich2012valid} introduce multivariate Mat\'ern models and provide necessary and sufficient conditions to allow the cross-covariance functions to have any number of components (processes) while allowing for different smoothnesses and rates of correlation decay for each component. We propose such a multivariate Mat\'ern construction to model our spatially varying intercepts, slopes, and residual SDs, so that each component is allowed its own spatial correlation structure.

In Section \ref{data}, we describe the motivating data. In Section \ref{methods}, we briefly review GPs and develop the SHREVE model. In Section \ref{results}, we apply the SHREVE model to GCC data and compare its performance to several nested models lacking visit effects or other model components. We give a concluding discussion in Section \ref{discussion}.


\section{Ganglion cell complex data} \label{data}

This section highlights data characteristics that motivate model development. We provide details on the imaging procedure and study subjects.

\subsection{Macular optical coherence tomography}

Macular OCT has emerged as a standard imaging modality to assess changes in retinal ganglion cells (RGCs) \citep{mohammadzadeh2020macular}. As glaucoma is characterized by progressive loss of RGCs, clinicians use macular OCT as a means to monitor changes in retinal thickness over time \citep{WEINREB20041711}. Macular GCC thickness, measured in microns ($\mu$m), has been shown to be more efficient for detecting structural loss regardless of glaucoma severity compared to measures of other macular layers \citep{mohammadzadeh2022ganglion}. Glaucomatous damage to the macular area, reflected in thinning of GCC, has been associated with VF loss \citep{mohammadzadeh2020longitudinal}. Visual field loss occurs when part(s) of the peripheral vision is (are) lost.

\subsection{Advanced Glaucoma Progression Study} \label{AGPS}

\begin{figure}[tp!]
	\centering
	\includegraphics[width=0.4\textwidth]{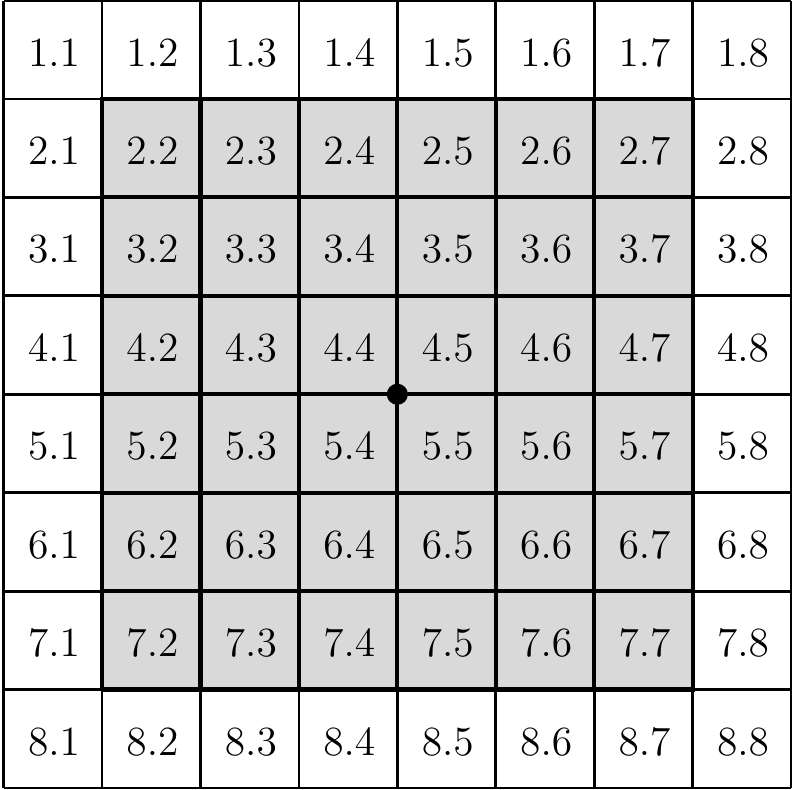}
	\caption{Visualization of the $8 \times 8$ grid of superpixels and labels from the Spectralis posterior pole algorithm. The inner 36 superpixels included in the analysis are shaded in gray and delineated with thicker lines. Superpixels are shown in right eye orientation where rows 1-4 are located in the \textit{superior hemiretina} and rows 5-8 are located in the \textit{inferior hemiretina}; the temple and nose are to the left and right, respectively. Superpixels labels are row number 1-8, a dot, then column number 1-8. The black dot indicates the foveal center for visual orientation.}
	\label{fig:grid}
\end{figure}

We analyze data from the AGPS \citep{mohammadzadeh2021estimating, mohammadzadeh2022ganglion, mohammadzadeh2022multivariate}, an ongoing longitudinal study at the University of California, Los Angeles. The study adhered to the tenets of the Declaration of Helsinki and conformed to Health Insurance Portability and Accountability Act policies. All patients provided written informed consent at the time of enrollment in the study. The data include GCC thickness measurements from 111 eyes with at least 4 OCT scans and a minimum of approximately 2 years of observed follow-up time, up to 4.25 years from baseline. Subjects returned approximately every 6 months for imaging using Spectralis OCT (Heidelberg Engineering, Heidelberg, Germany). This device acquires $30\degree \times 25\degree$ volume scans centered on the fovea, the center of the macula represented as a white dot in Figure \ref{fig:grid} and as a black dot in subsequent figures \citep{mohammadzadeh2020macular}. We used built-in software, the Glaucoma Module Premium Edition, to automatically segment macular layers of interest. GCC thickness is calculated by summing the thicknesses of the retinal nerve fiber layer, inner plexiform layer, and ganglion cell layer. The posterior pole algorithm of the Spectralis reports layer thickness averaged over pixels within a \textit{superpixel} with superpixels forming an 8 $\times$ 8 grid of locations, as shown in Figure \ref{fig:grid}. We display superpixels in right eye orientation with superpixels labeled as row number 1-8, a dot, then column number 1-8. Superpixels in rows 1-4 are located in the \textit{superior hemiretina} and rows 5-8 are located in the \textit{inferior hemiretina}; the temple and nose are to the left and right, respectively. Left eyes are mirror images of right eyes and are flipped left-right for presentation and analysis. Because there is substantial measurement noise in the outer ring of superpixels, rows 1 and 8 and columns 1 and 8 \citep{miraftabi2016local}, we analyze only the central 6 $\times$ 6 superpixels as shown in Figure \ref{fig:grid}. 

\begin{figure}[htp!]
	\centering
	\includegraphics[width=0.9\textwidth]{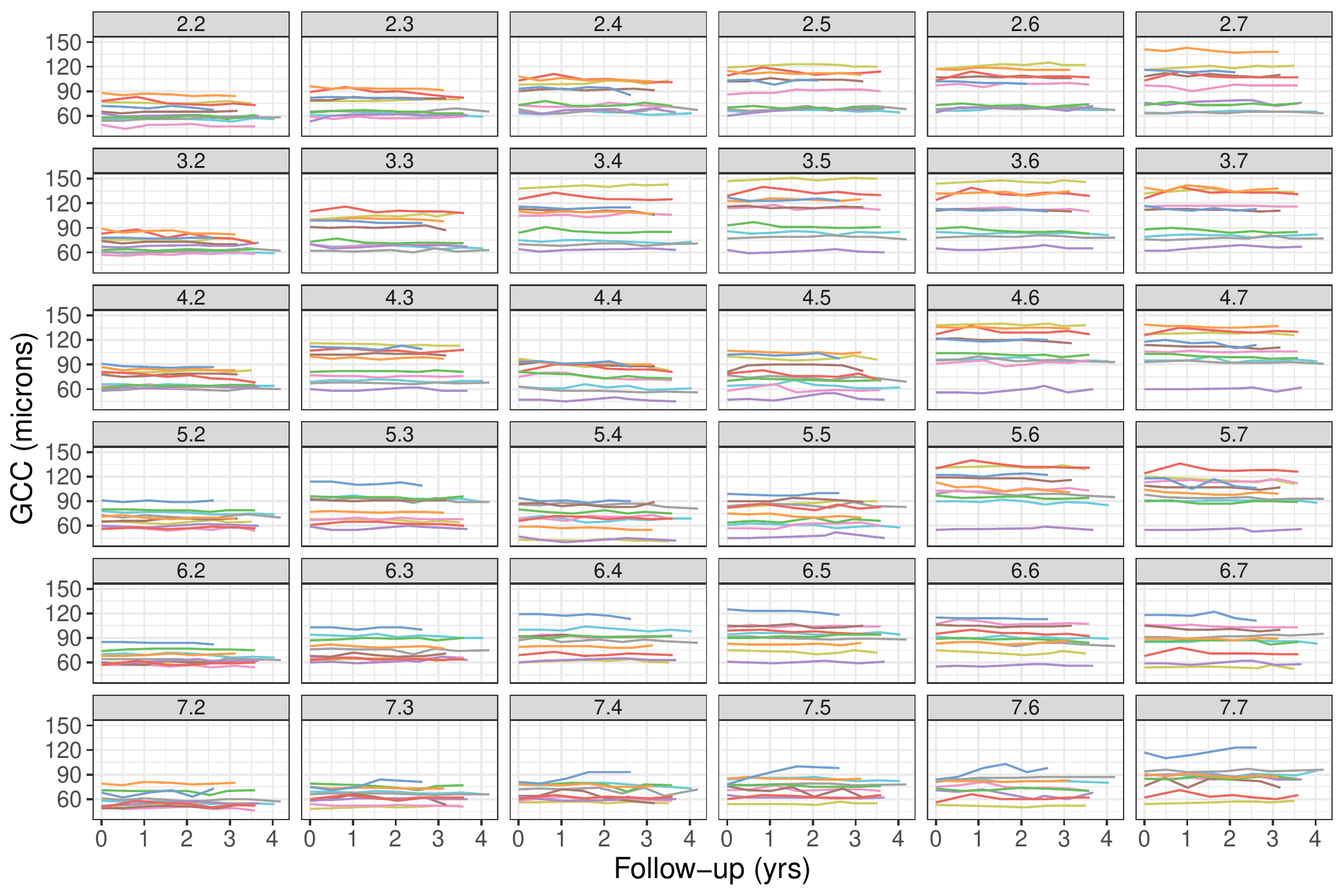}
	\caption{Profile plots of ganglion cell complex (GCC) thickness measurements for 10 subjects across 36 superpixels against follow-up time in years since baseline visit. Each color represents a different subject. These profiles illustrate the variability in baseline GCC thickness across the 10 subjects within superpixels, with a range within a superpixel of up to 84 $\mu$m. The average baseline thicknesses over subjects vary across superpixels, generally increasing from the temporal to nasal regions (left to right).}
	\label{fig:profile}
\end{figure}

\subsection{Data exploration}

Let observation $y_{ijk}$ be the GCC thickness measure in $\mu$m of subject $i = 1, \dots, n$ at visit $j = 1, \dots, J_i$, where $J_i$ is the number of visits for subject $i$, in superpixel $k = 1, \dots, K$ observed at time $t_{ij}$, with $t_{i1} = 0$ for all subjects. Location $\boldsymbol{s}_k = (\mbox{row}_k, \mbox{column}_k)$ denotes the spatial coordinates of superpixel $k$ in two-dimensional space. Initially, we remove any zero thickness values $y_{ijk} = 0$, which indicate errors of measurement. We define a profile for subject $i$ in superpixel $k$ as the sequence of observations ($t_{ij}, y_{ijk}$) from visits $j = 1, \dots, J_i$ and plot profiles of GCC thickness against time by connecting consecutive observations with line segments. For all subjects and superpixels, we plotted data in profile plots, which identified a number of outliers. We applied a semi-automated algorithm to identify pairs of consecutive points that have large differences in GCC thicknesses between the consecutive visits. For each subject and superpixel, we calculated the consecutive-visit absolute differences $| y_{ijk} - y_{i(j-1)k} |$ and the consecutive-visit centered-slopes $| y_{ijk} - y_{i(j-1)k} \ (t_{ij} - t_{i(j-1)}) + 0.5 |$, which were centered around $-0.5$ $\mu$m/year, the mean of slopes across all pairs of consecutive visits for all subjects and superpixels. We flagged pairs of observations ($y_{ijk}, y_{i(j-1)k}$) with absolute centered-slopes greater than 24 $\mu$m/year with absolute differences greater than 5 $\mu$m  as candidates for removal. We calculated the sum of absolute visit differences for each profile $\sum_{j=2}^{J_i} | y_{ij} - y_{i(j-1)k}|$ and removed the point that resulted in the largest reduction in the sum of absolute visit differences. For each profile, if two or more observations were identified as outliers, we removed all remaining observations as well.

Eyes enrolled in the AGPS had moderate to severe glaucoma, thus exhibit a range of glaucomatous damage. Figure \ref{fig:profile} shows profile plots after outlier removal of GCC thickness in $\mu$m against time in years since baseline visit for 10 subjects at all 36 superpixels. Baseline GCC varies across subjects within superpixels, with maximum differences in thicknesses between any of the AGPS subjects ranging from 40 to 100 $\mu$m across superpixels. From Figure \ref{fig:profile}, we note that intercepts are spatially correlated and repeated thickness measurements for each subject at each superpixel are highly correlated. The leftmost, temporal superpixels tend to have lower baseline thicknesses and smaller spread than rightmost, nasal superpixels and nasal superpixels show more variability both within and between subjects.

\begin{figure}[tp!]
	\centering
	\includegraphics[width=1\textwidth]{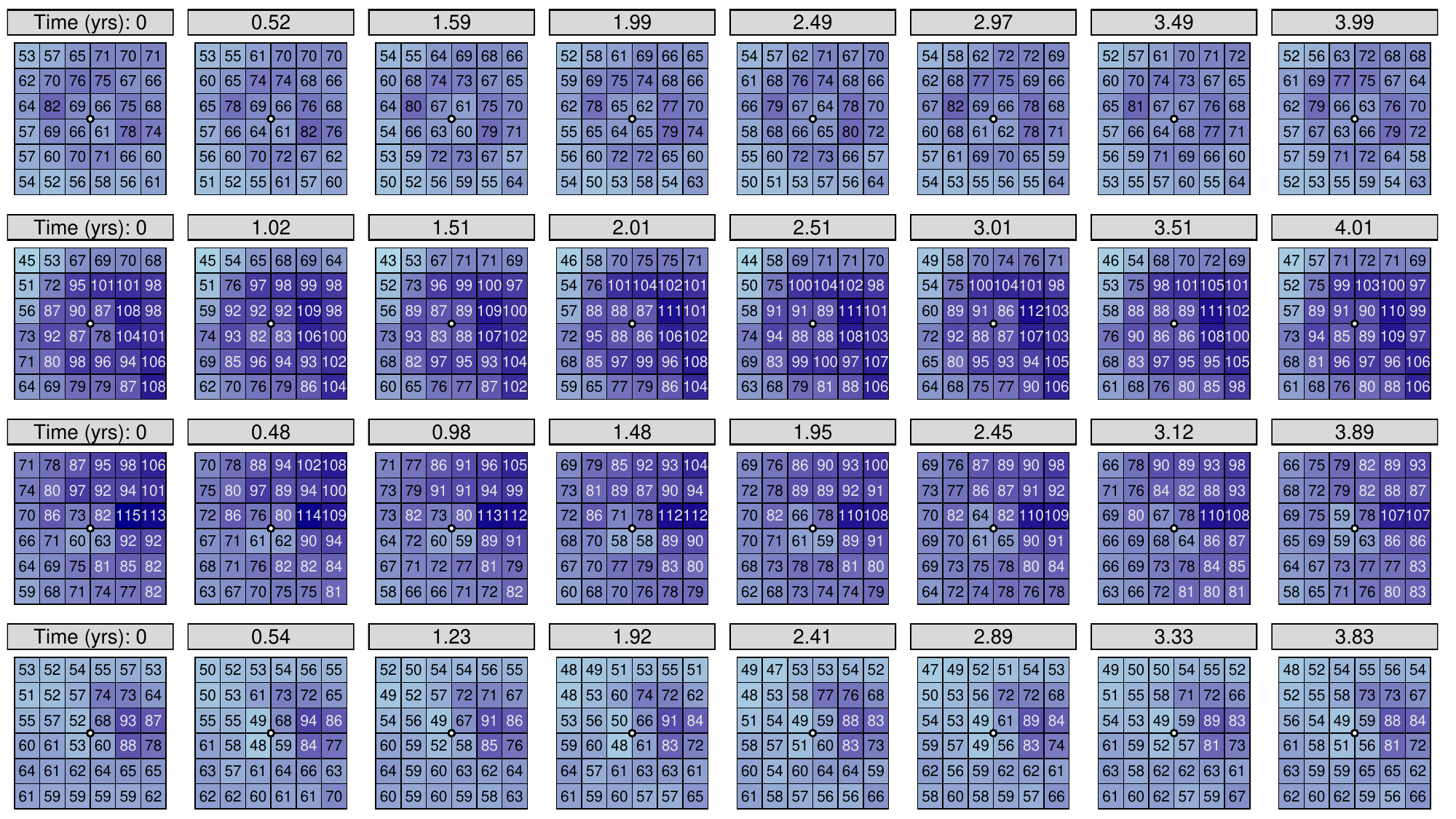}
	\caption{Heatmaps of ganglion cell complex (GCC) thickness measurements ($\mu$m) across 8 visits for 4 subjects for all 36 superpixels (top left 2.2 to bottom right 7.7). Each row is a different subject. The follow-up time of each visit is labeled at the top of each block. All maps share a common color scale for comparison. GCC measurements are highly correlated within subjects over time, illustrated by similar color patterns over time. The color patterns also highlight the  spatial correlation between locations. GCC measurements are highly variable across subjects, as seen by the difference in color shades. Over time, the third row subject has noticeable thinning in many superpixels while the other subjects are more stable in comparison.}
	\label{fig:gcc}
\end{figure}

\begin{figure}[tp!]
	\centering
	\includegraphics[width=1\textwidth]{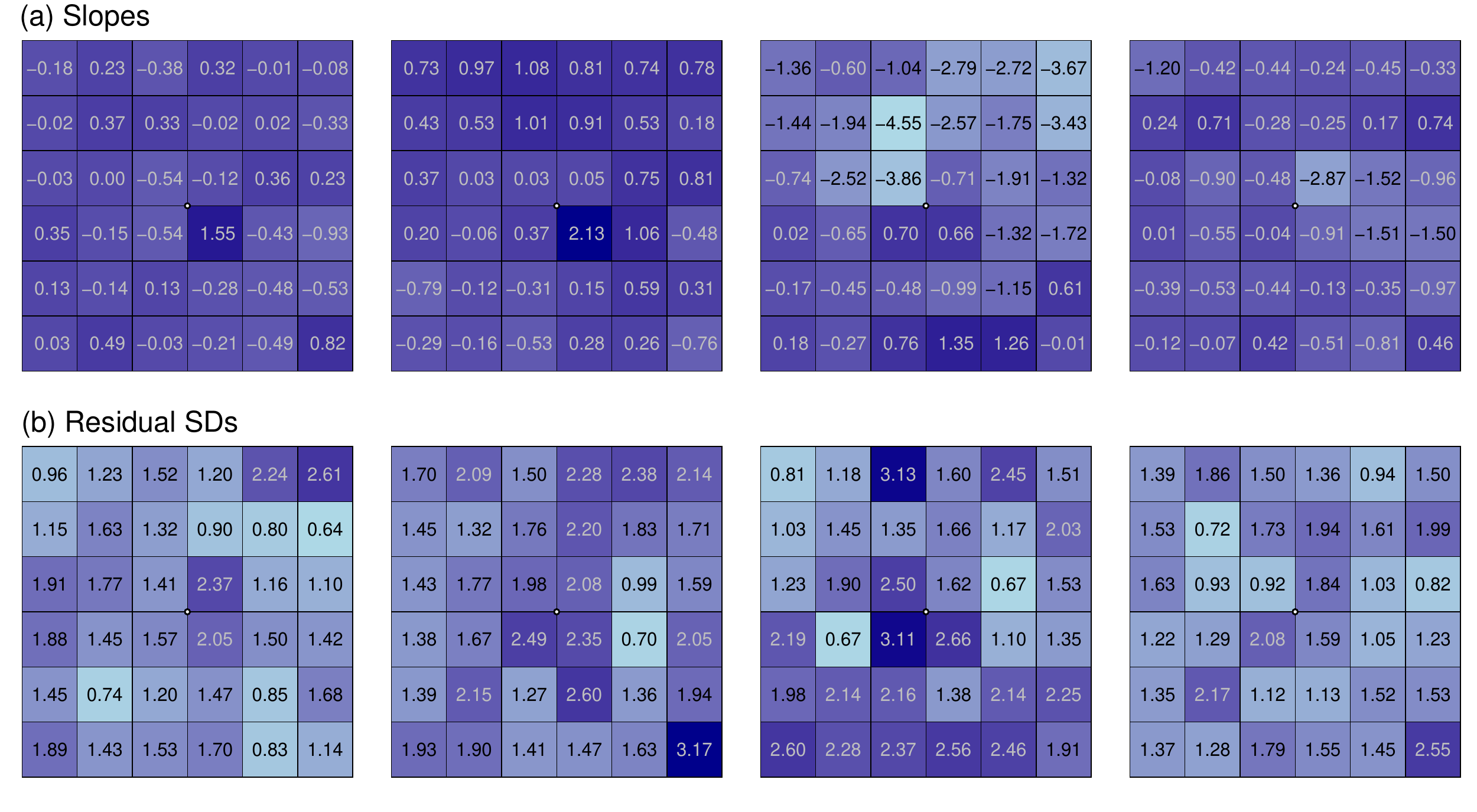}
	\caption{Heatmaps of (a) estimated slopes ($\mu$m/year) and (b) residual standard deviations (SD) ($\mu$m) for the same 4 subjects as in Figure \ref{fig:gcc} using simple linear regressions of ganglion cell complex (GCC) thicknesses on time since baseline in each superpixel. Each column is a different subject. Estimated slopes appear spatially correlated within subjects. Subject 3 has particularly steep negative slopes in the upper half of the eye, while Subjects 1 and 2 have more stable slopes across superpixels. The estimated residual SDs vary within subject by superpixel location. Subjects 1 and 4 have more uniform residual SDs across locations while Subjects 2 and 3 have some superpixels with much higher residual SDs.}
	\label{fig:slr}
\end{figure}

Figure \ref{fig:gcc} shows heatmaps of GCC measurements over time for four subjects. Each row represents a different subject and each block of $6 \times 6$ superpixels displays the GCC thicknesses observed in rows 2-7 and columns 2-7 at the labeled follow-up time above the block. The range of baseline thicknesses across superpixels varies across subjects, with the first subject's baseline values ranging between 53 and 82 $\mu$m, while the third subject's baseline values range between 59 and 115 $\mu$m. Changes in GCC thickness over time also differ between Subject 1 and Subject 3. Subject 3 has noticeable decrease in thickness, thinning over time in many superpixels (e.g., 2.7, 3.3, and 4.3), while Subject 1 is more stable over time. Within subjects, there is a range of baseline thicknesses and changes over time across superpixels. These data characteristics motivate the need to model spatially varying random intercepts and slopes. Analyzing longitudinal GCC data separately in each superpixel, \citet{mohammadzadeh2021estimating} show that models with subject-specific residual SDs perform better than models with fixed residual SDs. Figure \ref{fig:slr} shows heatmaps of estimated slopes (top) and residual SDs (bottom) from SLR of GCC thickness on time since baseline in each superpixel for the same four subjects as in Figure \ref{fig:gcc}, where each column is a different subject. Estimated slopes and residual SDs appear spatially correlated.

\begin{figure}[tp!]
	\centering
	\includegraphics[width=0.9\textwidth]{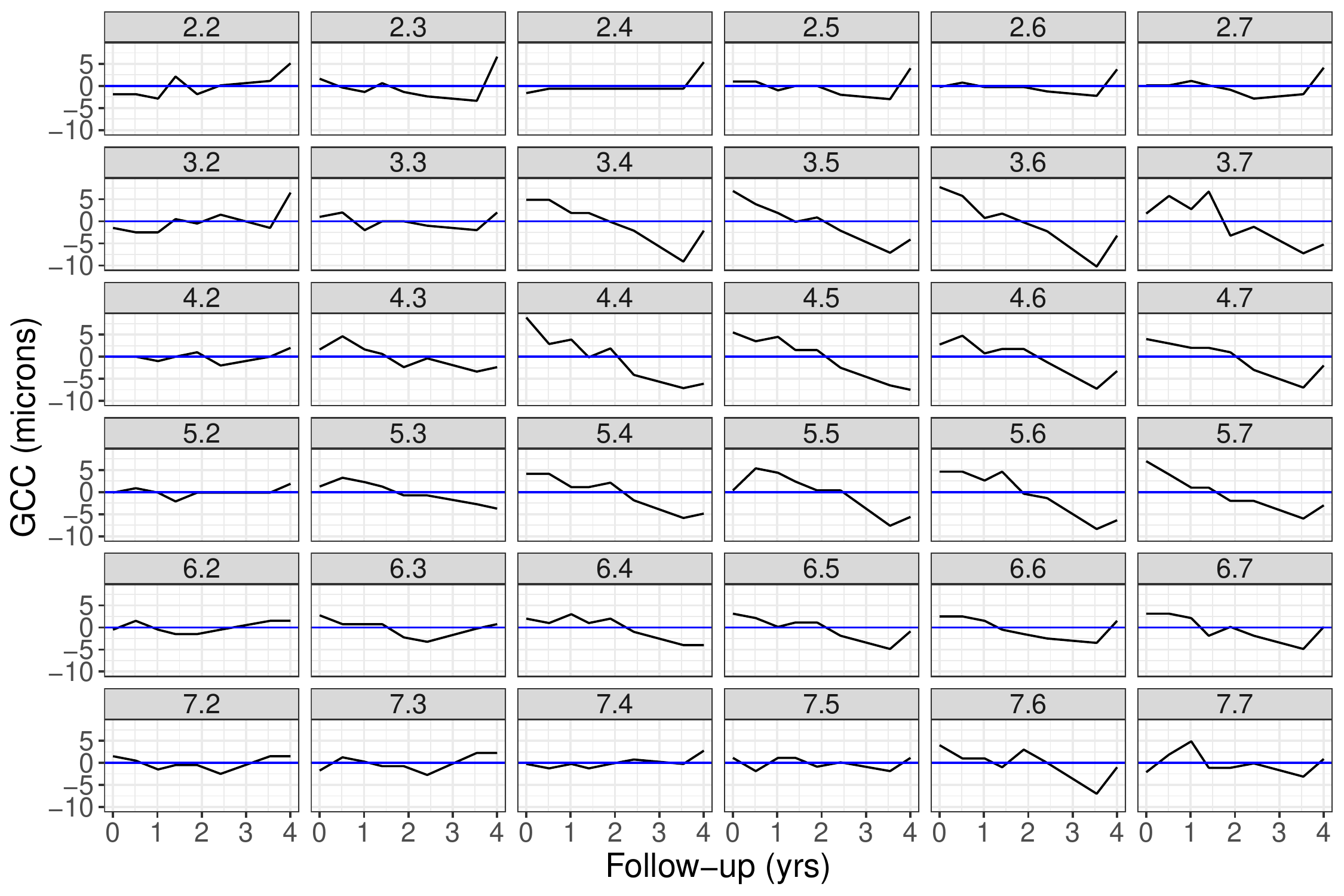}
	\caption{Empirical residual profile plots (superpixel mean subtracted from ganglion cell complex (GCC) thickness) for a single subject across 36 superpixels. There is an increase at four years for many superpixel locations suggesting visit-specific spatially correlated errors.}
	\label{fig:vis_eff}
\end{figure}

\citet{bryan2015global} model errors that affect all locations at a visit in glaucomatous VFs as global visit effects. Similar to VF data, we suspect there are spatially correlated errors in GCC measurements. We speculate these effects arise from the imaging process and segmentation errors that affect multiple locations. To better visualize these effects, we plot empirical residuals $y_{ijk} - \overbar{y}_{ik}$, where $\overbar{y}_{ik} = \sum y_{ijk} / J_i$. Empirical residual profile plots allow us to better see time trends within and across superpixels. Figure \ref{fig:vis_eff} provides an example of correlated errors across superpixels, where there is a noticeable increase at four years of follow-up. It is unlikely that such an increase is due to thickening of GCC, but rather due to errors in the imaging process or layer segmentation. Figure \ref{fig:vis_eff} shows  spatially correlated slopes noticeable in the region from superpixels 3.4 to 3.7 down to 6.4 to 6.7.

\subsection{Modeling goals}

We are interested in estimating individual rates of change at the superpixel level and predicting future GCC observations. To this end, we explicitly model the correlations between intercepts, slopes, and residual SDs at both the population and subject level. The intercepts are correlated with the magnitude of the slopes; as the baseline thickness increases, rates of change are faster \citep{rabiolo2020comparison}. Healthier eyes tend to have more thickness at baseline, with more potential for progression but also more opportunities for clinicians to intervene and prevent vision loss. Accounting for the relationships between measurement variability and either baseline thickness or slopes may help to better estimate the rates of progression and elucidate whether increased noise is associated with worsening disease. As glaucoma progresses, the ganglion cell and inner plexiform layers, two sublayers of GCC, show increased measurement variability especially as measures tend towards their floor \citep{miraftabi2016local}.


\section{Methods} \label{methods}

This section reviews the MGP priors we use to model the spatially varying visit effects and coefficients, constructs the SHREVE model, defines the priors, and introduces model comparison metrics.

\subsection{Gaussian processes} \label{method1}

A Gaussian spatial process \citep{williams2006gaussian, bogachev1998gaussian, banerjee2003hierarchical} is a stochastic process $\{z(\mathbf{s}): \mathbf{s} \in \mathbb{R}^d \}$ in which any finite collection of real-valued random variables $\{z(\mathbf{s}_1), \dots, z(\mathbf{s}_K)\}$ is distributed as multivariate normal for every set of $K \ge 1$ spatial locations $\mathbf{s}_1, \dots, \mathbf{s}_K \in \mathbb{R}^d$, for dimension $d \ge 1$; we work only with $d = 2$. We denote a GP as 
$$ z(\mathbf{s}) \sim \mbox{GP}(m(\mathbf{s}), C(\mathbf{s}, \mathbf{s}')), $$
with mean function $m(\mathbf{s}) = \mathbb{E}[z(\mathbf{s})]$ and covariance function $C(\mathbf{s}, \mathbf{s}') = \mbox{cov}[z(\mathbf{s}), z(\mathbf{s}')]$ for two locations $\mathbf{s}$ and $\mathbf{s}'$, which may be the same or distinct. The covariance function $C(\mathbf{s}, \mathbf{s}')$ models how similar outcomes $z(\mathbf{s})$ and $z(\mathbf{s}')$ are. We assume stationary and isotropic covariance functions $C(\mathbf{s}, \mathbf{s}')$. Stationarity means $C(\mathbf{s}, \mathbf{s}')$ depends only on the spatial separation vector $\textbf{s} - \textbf{s}'$ between points, and isotropy means $C(\mathbf{s}, \mathbf{s}')$ depends only on the distance between locations $h = \| \mathbf{s} - \mathbf{s}' \|$, where $\| \cdot \|$ is the Euclidean norm, i.e., $C(\mathbf{s}, \mathbf{s}') \equiv C(h)$.

We use Mat\'ern covariance functions of the form $\sigma^2 M(h| \nu, \ell)$, where $\sigma^2 > 0$ is the variance and $M(h| \nu, \ell)$ is the Mat\'ern correlation function \citep{Matern1986}
$$ M(h| \nu, \ell) = \frac{2^{1-\nu}}{\Gamma(\nu)}(\sqrt{2\nu} h / \ell)^\nu K_{\nu}(\sqrt{2\nu} h / \ell), $$
where $\nu >0$ is the smoothness parameter, $\ell > 0$ is the lengthscale, and $K_{\nu}$ is the modified Bessel function of the second kind of order $\nu$ \citep{abramowitz1964handbook}. In general, the process is $m$ times mean square differentiable if and only if $\nu > m$ \citep{williams2006gaussian}. The lengthscale parameter $\ell$ controls how quickly the correlation decays as a function of distance with larger $\ell$ indicating slower correlation decay.

\subsection{Multivariate Gaussian processes} \label{method2}

Let $\mathbf{z(\mathbf{s})} = (z_1(\mathbf{s}), \dots, z_P(\mathbf{s}))^T$ be a $P \times 1$ stochastic process where each component $z_p(\mathbf{s})$ for $p = 1, \dots, P$ is a scalar random variable at location $\mathbf{s}$. Then $\mathbf{z(\mathbf{s})}$ is an MGP if any random vector $(\mathbf{z}(\mathbf{s}_1)^T, \dots, \mathbf{z}(\mathbf{s}_K)^T)^T$ from any set of $K \ge 1$ locations $\mathbf{s}_1, \dots, \mathbf{s}_K$ has a multivariate normal distribution. The MGP is an extension of the univariate GP where the random variables $\mathbf{z(\mathbf{s})}$ are vector-valued. We denote an MGP as
$$ \mathbf{z}(\mathbf{s}) \sim \mbox{MGP}(\textbf{m}(\mathbf{s}), \textbf{C}(\mathbf{s}, \mathbf{s}')), $$
with $P \times 1$ mean vector $\textbf{m}(\mathbf{s})$ and $P \times P$ cross-covariance matrix function $\textbf{C}(\mathbf{s}, \mathbf{s}') =  \mbox{cov}[\mathbf{z}(\mathbf{s}), \mathbf{z}(\mathbf{s}')] = \{C_{pq}(\mathbf{s}, \mathbf{s}')\}_{p,q = 1}^P$. Functions $C_{pq}(\mathbf{s}, \mathbf{s}') = \mbox{cov}[z_p(\mathbf{s}), z_q(\mathbf{s}')]$, for $p, q = 1, \dots, P$, are called marginal covariance functions when $p = q$ and cross-covariance functions when $p \ne q$.

We want to allow each marginal process to have its own spatial correlation function. Each marginal covariance function $C_{pp}$ is modeled with a Mat\'ern correlation function, $C_{pp}(h) = \sigma_{pp}^2 M(h| \nu_{pp}, \ell_{pp})$, for $p = 1, \dots, P$, with variance parameter $\sigma_{pp}^2 >0$, smoothness parameter $\nu_{pp}$, and lengthscale parameter $\ell_{pp}$. We model each cross-covariance function $C_{pq}$ with a Mat\'ern correlation function, $C_{pq}(h) = \sigma_{pq} M(h| \nu_{pq}, \ell_{pq})$, for $1 \le p \ne q \le P$, with covariance parameter $\sigma_{pq}$, smoothness parameter $\nu_{pq}$, and lengthscale parameter $\ell_{pq}$. We assume marginal covariance $C_{pp}$ and cross-covariance $C_{pq}$ functions to be Mat\'ern following sufficient conditions on parameters $\nu_{pp}$, $\nu_{pq}$, $\ell_p$, $\ell_{pq}$, $\sigma_{pp}$, and $\sigma_{pq}$ that result in a nonnegative definite cross-covariance function \citep{apanasovich2012valid}. We use the simplest parameterization, where no additional parameters beyond $\sigma_{pp}^2$, $\nu_{pp}$, and $\ell_{pp}$ are required to model the smoothness and lengthscale parameters for the cross-covariances. The cross-covariance function $\textbf{C}(\mathbf{s}, \mathbf{s}')$ is nonnegative definite when
\begin{align}
	\nu_{pq}(\nu_{pp}, \nu_{qq}) &= \frac{\nu_{pp} + \nu_{qq}}{2}, \nonumber \\
	\ell_{pq}(\ell_p, \ell_q) &= \sqrt{\frac{2}{\ell_p^{-2} + \ell_q^{-2}}}, \label{eq:2}\\
	\sigma_{pq}(\nu_{pp}, \nu_{qq}, \ell_p, \ell_q, \sigma_{pp}, \sigma_{qq}, R_{pq}) &= \sigma_{pp} \sigma_{qq}\frac{\ell_{pq}(\ell_p, \ell_q)}{\sqrt{\ell_p \ell_q}} \frac{\Gamma(\nu_{pq}(\nu_{pp}, \nu_{qq}))}{\Gamma^{1/2}(\nu_{pp}) + \Gamma^{1/2}(\nu_{qq})} R_{pq}, \label{eq:3}
\end{align}
where $\mathbf{R} = \{R_{pq} \}$ is a nonnegative definite $P \times P$ correlation matrix with diagonal elements equal to 1 and nondiagonal elements in the closed interval [-1, 1]. The cross-correlation $\rho_{pq} = \sigma_{pq} / \sigma_{pp} \sigma_{qq} = \mbox{corr}(z_p(\mathbf{s}), z_q(\mathbf{s}))$ is the correlation between $z_p(\mathbf{s})$ and $z_q(\mathbf{s})$.

\subsection{Model specification for a spatially varying hierarchical random effects with visit effects model}

\begin{figure}[t]
	\centering
	\includegraphics[width=0.75\textwidth]{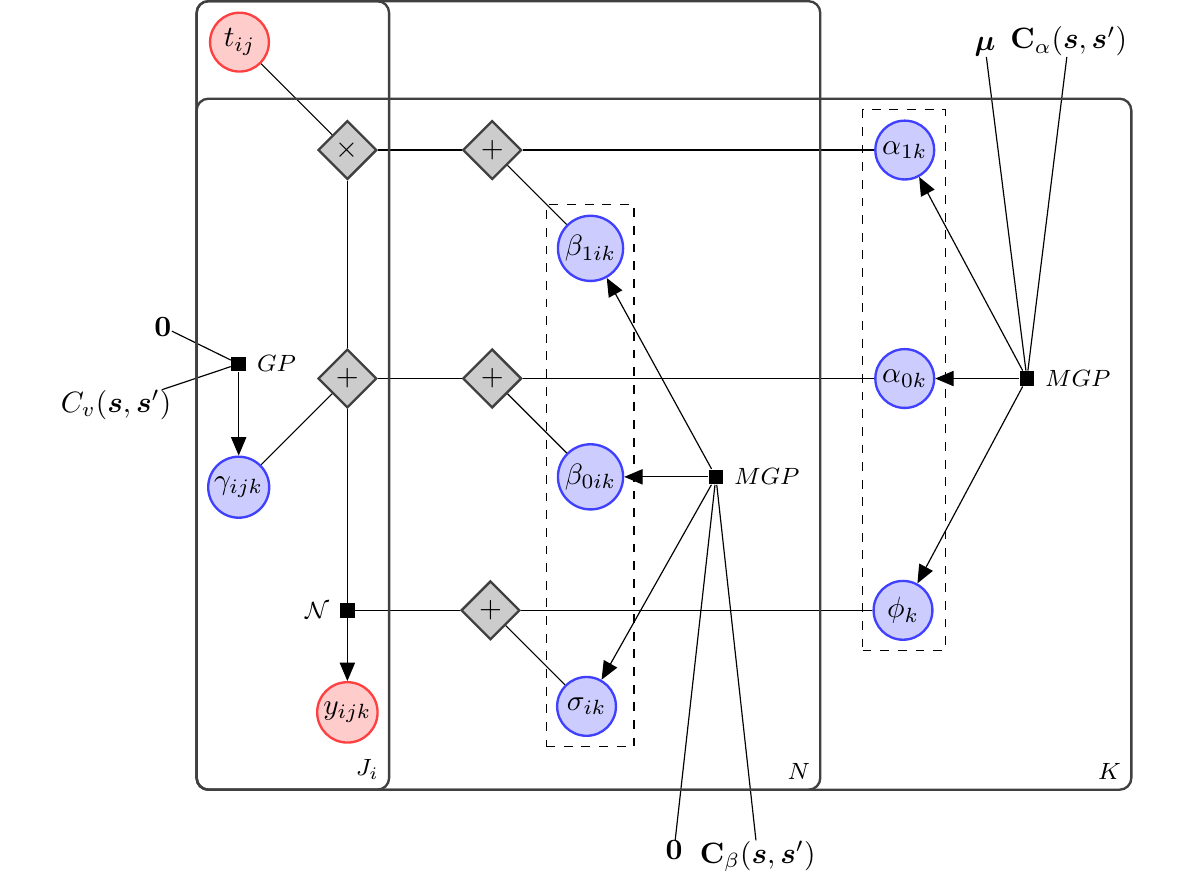}
	\caption{Plate diagram of the proposed model. Blue nodes are latent variables, red nodes are observed variables, gray nodes are deterministic nodes, GP stands for \emph{Gaussian process}, and MGP stands for \emph{multivariate Gaussian process}. Plates are used to group variables repeated together over subjects, time, and space, where $i = 1, \dots, N$ indexes subjects, $j = 1, \dots, J_i$ indexes subject $i$'s visits, and $k = 1, \dots, K$ indexes superpixel locations.}
	\label{fig:plate}
\end{figure}

The proposed SHREVE model allows random intercepts, slopes, and log residual SDs to be correlated within and across locations while accounting for within-subject variability and spatially correlated visit-specific errors. For ease of notation, we specify the model assuming no missing data but note that complete data is not a requirement. We model $y_{ijk}$ as
\begin{align*}
	y_{ijk} &= \alpha_{0k} + \alpha_{1k} t_{ij} + \beta_{0ik} + \beta_{1ik} t_{ij} + \gamma_{ijk} + \epsilon_{ijk} \\
	\epsilon_{ijk} | \tau^2_{ik} &\sim \mbox{N}(0, \tau^2_{ik}), \nonumber \\
	\log \tau_{ik} &= \phi_k + \sigma_{ik}, \nonumber
\end{align*}
where $\alpha_{0k}$, $\alpha_{1k}$, and $\phi_k$ are the superpixel $k$ population-level intercept, slope, and log residual SD processes, respectively, $\beta_{0ik}$, $\beta_{1ik}$, and $\sigma_{ik}$ are subject-specific intercept, slope, and log residual SD processes, respectively, in superpixel $k$ and $\gamma_{ijk}$ is the visit effect process at location $\mathbf{s}_k$ for subject $i$ visit $j$. Figure \ref{fig:plate} presents the model graphically. 

Let $\boldsymbol{\alpha}_k = (\alpha_{0k}, \alpha_{1k}, \phi_k)^T$ denote the population-level (PL) multivariate spatial process, which we model with MGP $\boldsymbol{\alpha}_k | \boldsymbol{\mu}, \boldsymbol{\theta}_\alpha  \sim \mbox{MGP}(\boldsymbol{\mu}, \mathbf{C}_\alpha(\mathbf{s}_k, \mathbf{s}_{k'}))$, with mean vector $\boldsymbol{\mu} = (\mu_0, \mu_1, \mu_{\phi})^T$ and PL cross-covariance matrix function $\mathbf{C}_\alpha(\mathbf{s}_k, \mathbf{s}_{k'})$ with hyperparameters $\boldsymbol{\theta}_\alpha = \{ \sigma_{\alpha, pp}, \nu_{\alpha, p}, \ell_{\alpha, p}, \mathbf{R}_{\alpha}, p \in \{1, 2, 3 \} \}$. The parameters $\mu_0$, $\mu_1$, and $\mu_{\phi}$ are the global grand mean intercept, slope, and log residual SD, respectively. PL marginal covariance functions $C_{\alpha, pp}(\mathbf{s}_k, \mathbf{s}_{k'}) = \sigma_{\alpha, pp}^2 M(h |\nu_{\alpha, p}, \ell_{\alpha, p})$, for $p = 1, ..., 3$ have PL marginal variances $\sigma_{\alpha, pp}^2$, PL smoothness parameters $\nu_{\alpha, p}$, and PL lengthscales $\ell_{\alpha, p}$. PL cross covariance functions $C_{\alpha, pq}(\mathbf{s}_k, \mathbf{s}_{k'}) = \sigma_{\alpha, pq} M(h|\nu_{\alpha, pq}, \ell_{\alpha, pq})$ have covariance parameters $\sigma_{\alpha, pq}$ between processes $p$ and $q$, smoothness parameters $\nu_{\alpha, pq}$, and lengthscales $\ell_{\alpha, pq}$. Here $h = \| \mathbf{s}_k - \mathbf{s}_{k'} \|$ is the distance between two superpixel locations, $\sigma_{\alpha, pq} \equiv \sigma_{pq}(\nu_{\alpha,p}, \nu_{\alpha,q}, \ell_{\alpha,p}, \ell_{\alpha,q}, \sigma_{\alpha,pp}, \sigma_{\alpha,qq}, R_{\alpha,pq})$ is a function of $\sigma_{\alpha, pp}$ and $\sigma_{\alpha, qq}$ as defined in (\ref{eq:3}), and $\ell_{\alpha, pq} \equiv \ell_{pq}(\ell_{\alpha,p}, \ell_{\alpha,q})$ is a function of $\ell_{\alpha, p}$ and $\ell_{\alpha, q}$ as in (\ref{eq:2}). The $3 \times 3$ cross-correlation matrix $\mathbf{R}_{\alpha}$ is an unknown symmetric matrix with 1's on the diagonal and with $(p,q)$th element the correlation parameter $R_{\alpha,pq} = R_{\alpha,qp}$.

\sloppy Similarly, we model random effects (RE) $\boldsymbol{\beta}_{ik} = (\beta_{0ik}, \beta_{1ik}, \sigma_{ik})^T$ as $\boldsymbol{\beta}_{ik} | \boldsymbol{\theta}_\beta \sim \mbox{MGP}(\boldsymbol{0}, \mathbf{C}_\beta(\mathbf{s}_k, \mathbf{s}_{k'}))$, with mean vector $\boldsymbol{0}$ and cross-covariance matrix function $\mathbf{C}_\beta(\mathbf{s}_k, \mathbf{s}_{k'})$ with hyperparameters $\boldsymbol{\theta}_\beta = \{ \sigma_{\beta, pp}, \nu_{\beta, p}, \ell_{\beta, p}, \mathbf{R}_{\beta}, p \in \{1, 2, 3 \} \}$. RE marginal covariance functions $C_{\beta, pp}(\mathbf{s}_k, \mathbf{s}_{k'}) = \sigma_{\beta, pp}^2 M(h |\nu_{\beta, p}, \ell_{\beta, p})$ for $p = 1, ..., 3$ have RE marginal variances $\sigma_{\beta, pp}^2$, smoothness parameters $\nu_{\beta, p}$, and lengthscales $\ell_{\beta, p}$. RE cross-covariance functions $C_{\beta, pq}(\mathbf{s}_k, \mathbf{s}_{k'}) = \sigma_{\beta, pq} M(\mathbf{h}|\nu_{\beta, pq}, \ell_{\beta, pq})$ have RE covariance parameters $\sigma_{\beta, pq} \equiv \sigma_{pq}(\nu_{\beta,p}, \nu_{\beta,q}, \ell_{\beta,p}, \ell_{\beta,q}, \sigma_{\beta,pp}, \sigma_{\beta,qq}, R_{\beta,pq})$, lengthscales $\ell_{\beta, pq} \equiv \ell_{pq}(\ell_{\beta,p}, \ell_{\beta,q})$, and unknown cross-correlation matrix $\mathbf{R}_{\beta}$ as defined in (\ref{eq:2}) and (\ref{eq:3}). We model the spatially varying visit effects $\gamma_{ijk}$ with mean 0 GPs $\gamma_{ijk} | \sigma_{v},\nu_{v}, \ell_{v} \sim \mbox{GP}(0, C_v(\mathbf{s}_k, \mathbf{s}_{k'}))$, with visit effects covariance function $C_v(\mathbf{s}_k, \mathbf{s}_{k'}) = \sigma^2_{v} M(\mathbf{h}|\nu_{v}, \ell_{v})$.

\subsection{Priors}

We use weakly informative priors to keep inferences within a reasonable range and allow computations to proceed satisfactorily. The closest two superpixels can be is 1 unit apart, and the largest separation is $\sqrt{(7-2)^2 + (7-2)^2} \approx 7$ units. We expect lengthscales to plausibly fall in this range. At the same time, we wish to avoid infinitesimal lengthscales. We assign independent and identical inverse gamma priors on all MGP lengthscale parameters $\ell_{\alpha, 1}$, $\ell_{\alpha, 2}$, $\ell_{\alpha, 3}$, $\ell_{\beta, 1}$, $\ell_{\beta, 2}$, $\ell_{\beta, 3}$, $\ell_{v} \sim IG(2.25, 2.5)$ with mean 2 and SD 4. For the MGP SD parameters, we wish to avoid flat priors that could pull the posterior towards extreme values. We assign truncated-normal priors on all MGP SD parameters $\sigma_{\alpha, 11}, \sigma_{\beta, 11} \sim N^+(0, 10^2)$, $\sigma_{\alpha, 22}, \sigma_{\alpha, 33}, \sigma_{\beta, 22}, \sigma_{\beta, 33}, \sigma_v \sim N^+(0, 2.5^2)$, where $N^+(a, b)$ is a normal distribution restricted to the positive real line with mean $a$ and variance $b$. We assign independent normal priors on the global effects $\mu_0 \sim N(73, 15^2)$, $\mu_1 \sim N(-0.3, 0.3^2)$, $\mu_{\phi} \sim N(0.7, 0.3^2)$. For the correlation matrices $\mathbf{R}_\alpha$ and $\mathbf{R}_\beta$, we assign marginally uniform priors on the individual correlations derived from the inverse Wishart distribution with $3 \times 3$ identity matrix scale matrix parameter and four degrees of freedom $IW(\mathbf{I}_3, 4)$ \citep{barnard2000modeling}. When $\boldsymbol{\Sigma}$ has a standard inverse-Wishart distribution, we can decompose $\boldsymbol{\Sigma} =\mathbf{SRS}$ in terms of the diagonal standard deviation matrix $\mathbf{S}$ and correlation matrix $\mathbf{R}$ to obtain the prior for the correlation matrices. We set all MGP smoothness parameters $\nu_{\alpha, 1}$, $\nu_{\alpha, 2}$, $\nu_{\alpha, 3}$, $\nu_{\beta, 1}$, $\nu_{\beta, 2}$, $\nu_{\beta, 3}$, $\nu_{v} = \tfrac{1}{2}$ since we obtain measurements from a coarse grid of superpixel locations and expect the processes to be rough. When $\nu = \frac{1}{2}$, the Matern correlation function reduces to the popular exponential kernel $M(\mathbf{h}| \frac{1}{2}, \ell) = \exp(-\| \mathbf{h} \| / \ell)$.

\subsection{Computation and inference}  \label{comp}

 For data analysis and visualization, we use the \textsc{R} programming language \citep{rlanguage} and \textsc{ggplot2} \citep{ggplot}. We use Markov Chain Monte Carlo (MCMC) methods \citep{metropolis1953equation, robert2005-kh} implemented in \textsc{nimble} v0.13.0 \citep{nimble}. We specify the model at the observation level and omit observations removed in the data cleaning step. To sample from the posteriors, we use Gibbs sampling and update specific parameters using the automated factor slice sampler or Metropolis-Hastings sampler within Gibbs. We update the global effects $\mu_0$, $\mu_1$ and $\mu_{\phi}$ using scalar Metropolis-Hastings random walk samplers; the visit effect GP lengthscale $\ell_{\nu}$ and subject-level residual SD GP SD parameter $\sigma_{\beta, 33}$ together using the automated factor slice sampler \citep{tibbits2014automated}; the subject-level random effects $\beta_{0ik}$, $\beta_{1ik}$, and $\sigma_{ik}$ and visit effects $\gamma_{ijk}$ using multivariate Metropolis-Hastings random walk samplers in sub-blocks. We tested various schemes for sampling sub-blocks of the subject-level random effects and visit effects to improve sampling efficiency \citep{risser2020bayesian}. We jointly sample subject-level intercepts, slopes, and the first visit effect in sub-blocks of size 3. We separately sample the subject-level residual SDs in sub-blocks of size 6 and the remaining visit effects in sub-blocks of size 3. Each pair of SD and lengthscale parameters from MGPs and GPs were sampled together (e.g., $(\sigma_{\alpha, 11}, \ell_{\alpha, 1})$) except for the subject-level residual SDs and visit effects where opposites were paired together $(\sigma_{\beta, 33}, \ell_{\nu})$ and $(\sigma_{\nu}, \ell_{\beta, 3})$. We run all models with 9 chains of 250,000 iterations after a burn-in of 30,000, a thin of 100 for a total of 19,800 posterior samples. Following \citet{vehtari2021rank}'s recommendation for assessing convergence, the bulk and tail effective sample sizes were all greater than 100 per chain and the potential scale reduction factor $\widehat{R}$ were all less than 1.01. Visual assessment of model convergence show satisfactory results. We show efficiency per iteration plots of the 7 parameters with the largest $\widehat{R}$ in Appendix Figure \ref{fig:ess} and summarize convergence diagnostics in Appendix Table \ref{tab:convergence}.

\subsection{Model comparison}

We fit the SHREVE model to the AGPS data and compare model fit of the SHREVE model to 7 nested models and to SLR fit separately for each subject and superpixel location. The 7 submodels were SHREVE omitting (a) the population-level residual SD process $\phi_k$, (b) the subject-specific residual SD process $\sigma_{ik}$, (c) the spatially varying visit effects $\gamma_{ijk}$, and all combinations (ab), (ac), (bc), and (abc). We call the SHREVE model without visit effects the spatially varying hierarchical random effects (SHRE) model. For SLR, we run a separate model for each eye and superpixel using flat priors with results equivalent to classical least squares.

We compare models with the Watanabe-Akaike (or widely applicable) information criterion (WAIC) \citep{watanabe2010asymptotic, Gelman2013-ht} and approximate leave-one-out cross-validation (LOO) using Pareto Smoothed Importance Sampling \citep{vehtari2017practical}. We report WAIC
$$
\mbox{WAIC} = -2 \left[ \sum_{i=1}^{n} \sum_{j=1}^{J_i} \sum_{k=1}^{K} \log \left( \frac{1}{S}\sum_{s=1}^{S} p(y_{ijk}|\theta^s) \right) - \sum_{i=1}^{n} \sum_{j=1}^{J_i} \sum_{k=1}^{K} V_{s = 1}^S \left( \log p(y_{ijk}|\theta^s) \right) \right]
$$
summing over all data points $y_{ijk}$, where $p(y_{ijk}|\theta)$ is the pointwise predictive density, $\theta$ are the model parameters, superscript $s$ denotes parameters drawn at the $s$th iteration for $s = 1, \dots, S$ posterior samples, and $V_{s = 1}^S$ denotes the sample variance over $S$ posterior samples. We report approximate LOO
$$
\mbox{LOO} = -2 \sum_{i=1}^{n} \sum_{j=1}^{J_i} \sum_{k=1}^{K} \log \left(\frac{ \sum_{s=1} w_{ijk}^s p(y_{ijk}|\theta^s)}{ \sum_{s=1} w_{ijk}^s } \right)
$$
where $w_{ijk}^s$, $s = 1, \dots, S$ is a vector of importance weights for data point $y_{ijk}$ at iteration $s$ and $w_{ijk}^s = \left(p(y_{ijk}|\theta^s)\right)^{-1}$ except for extreme weights. Approximate LOO estimates the out-of-sample predictive accuracy of the model \citep{stone1977asymptotic}. Lower WAIC and LOO indicate better fit. 

To assess predictive accuracy of the proposed model, we compare models on mean squared prediction error 
$$
\mbox{MSPE} = \frac{\sum_{s=1}^{S} \sum_{i=1}^{n} \sum_{k \in \mathcal{K}_i} (y_{i J_{i} k} - \hat{y}_{i J_{i} k}^{s})^2}{SN_{pred}}
$$ 
for $s = 1, \dots, S$ posterior MCMC samples, $i = 1, \dots, n$ subjects, $k \in \mathcal{K}_i$ held out superpixels for subject $i$, held out observations $y_{iJ_{i}k}$, and predicted observations for each posterior sample $\hat{y}^s_{iJ_{i}k}$, of $N_{pred}$ total held out observations after fitting the models. We randomly sample and hold out 7 observations $y_{iJ_{i}k}$, or approximately 20\%, at the last visit for each of 110 subjects and 6 observations for one subject because they only had 32 observations available at the last visit, for a total of $N_{pred} = 111 \times 7 - 1 = 776$ observations, and fit models with the remaining observations. Not all observations are available at all superpixels because we remove some observations in the data cleaning step. For the SHREVE models, we define a predicted observation at each posterior sample $s$ as 
\begin{align}
	\hat{y}^s_{iJ_{i}k} = \alpha^s_{0k} + \alpha^s_{1k} t_{iJ_{i}} + \beta^s_{0ik} + \beta^s_{1ik} t_{iJ_{i}} + \gamma_{iJ_{i}k}, \label{eq:4}
\end{align}
where $t_{iJ_{i}}$ is the time observed and $\gamma_{iJ_{i}k}$ is the visit effect for the held out observation at the $i$th subject's last visit. For the SHRE models, there is no $\gamma_{iJ_{i}k}$ visit effect term in (\ref{eq:4}). 


\section{Advanced Glaucoma Progression Study} \label{results}


\begin{table}[t]
	\rowcolors{1}{}{gray!25}
	\centering
	\caption{Model fit comparison with widely applicable information criterion (WAIC), approximate leave-one-out cross-validation with Pareto Smoothed Importance Sampling (LOO), and mean squared prediction error (MSPE) of predictions. For predictions, we hold out 7 randomly sampled observations $y_{iJ_{i}k}$ at the last visit of each of 110 AGPS subjects and 6 observations from one subject. Models with visit effects perform better than models without visit effects. SLR performs noticeably worse compared to the hierarchical models. The smallest WAIC, LOO, and MSPE values are bolded.}
	\label{tab:waic}
	\begin{tabular}{c>{\centering} p{0.05\textwidth}>{\centering} p{0.07\textwidth}> {\centering\arraybackslash} p{0.12\textwidth} >{\centering} p{0.12\textwidth} c >{\centering} p{0.08\textwidth} >{\centering\arraybackslash} p{0.06\textwidth} }
		\toprule
		\textbf{Model} & \textbf{Joint Model} & \textbf{Visit Effects} & \textbf{Superpixel Residual SD} & \textbf{Subject Residual SD} & \textbf{WAIC} & \textbf{LOO} & \textbf{MSPE} ($\mu\mbox{m}^2$) \\
		\midrule
		SHREVE & \cmark & \cmark & \cmark & \cmark & \textbf{107,581.6} & \textbf{113,323.1} & 6.6 \\
		SHREVE-(a) & \cmark & \cmark & \xmark & \cmark & 108,002.2 & 113,560.7 & \textbf{6.5} \\
		SHREVE-(b) & \cmark & \cmark & \cmark & \xmark & 110,992.3 & 116,978.1 & 6.8 \\
		SHREVE-(ab) & \cmark & \cmark & \xmark & \xmark & 113,238.3 & 118,647.5 & 6.9 \\
		SHRE & \cmark & \xmark & \cmark & \cmark & 124,389.5 & 125,304.7 & 7.2 \\
		SHRE-(a) & \cmark & \xmark & \xmark & \cmark & 124,468.8 & 125,461.2 & 7.1 \\
		SHRE-(b) & \cmark & \xmark & \cmark & \xmark & 129,353.2 & 129,877.3 & 7.5 \\
		SHRE-(ab) & \cmark & \xmark & \xmark & \xmark & 130,188.4 & 130,732.1 & 7.5 \\
		SLR & \xmark & \xmark & \xmark & \cmark & 128,870.2 & 132,916.3 & 39.7 \\
		\bottomrule
	\end{tabular}
\end{table}

\begin{figure}[t!]
	\centering
	\includegraphics[width=0.9\textwidth]{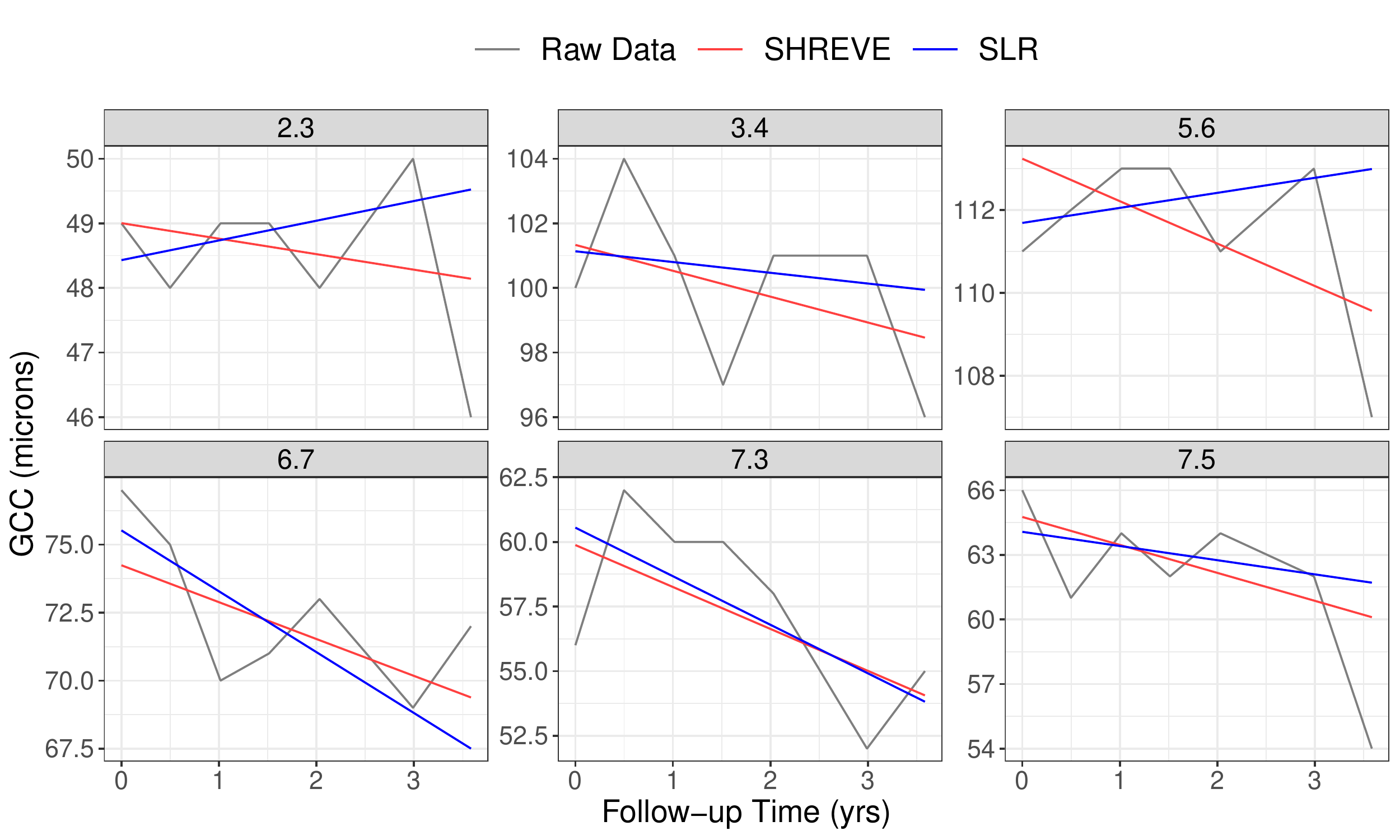}
	\caption{Comparison of predicted observations and model fit from the SHREVE model and simple linear regression (SLR) after holding out the last observation at 3.6 years follow-up of this subject. The gray line plots the raw data, the red line is the posterior mean fitted line from the SHREVE model without adding in the visit effects, and the blue line shows the fitted line from SLR. The SHREVE model is able to better estimate slopes and predict the last observation in noisy superpixels like 2.3 and 5.6 than SLR.}
	\label{fig:predicted}
\end{figure}

After identifying and removing approximately 0.5\% of the data as outliers, we analyze 29,179 observations from 111 subjects over 36 superpixels. Table \ref{tab:waic} gives the WAIC, LOO, and MSPE of models considered. The SHREVE model has the lowest WAIC and LOO. Comparing pairs of SHREVE and SHRE models with and without the (a) population-level residual SD process and (b) subject-level residual SD process, omitting (a) increases WAIC (LOO) by up to 421 (238) while omitting (b) increases WAIC (LOO) by up to 4,964 (4,573). Omitting visit effects increases WAIC (LOO) by up to 18,361 (12,899). SLR has lower WAIC than the two SHRE models without (b), but SLR still has higher LOO. Having subject-specific residual SDs is more important for models without a visit effect component, as the difference in WAIC (LOO) between SHRE and SHRE-(b) is larger by 1553 (918) than the difference between SHREVE and SHREVE-(b). For predictions, the MSPE for SLR is 6.0 times that of the SHREVE model (39.7 vs. 6.6 $\mu\mbox{m}^2$) and 5.5 times that of the SHRE model (39.7 vs. 7.2 $\mu\mbox{m}^2$). Among the hierarchical models, the biggest distinction in MSPE is between models with and without visit effects. Comparing pairs of SHREVE and SHRE models, omitting the subject-level residual SD process consistently increases the MSPE, while omitting the population-level residual SD process has a negligible effect on MSPE. Figure \ref{fig:predicted} plots profiles and posterior mean fitted lines from the SHREVE model and SLR for one subject for superpixels that had the last (7th) observation held out. The SHREVE model better estimates slopes for noisy superpixels like 2.3 and 5.6. All predictions of the last visit in the 6 superpixels by the SHREVE model are closer to the GCC observed at $t_{ij} = 3.6$ than those by SLR.

\begin{table}[t]
	\centering
	\caption{Posterior mean and 95\% credible interval (CrI) for global parameters and subject-level multivariate Gaussian process (MGP) parameters comparing the SHREVE and SHRE models.}
	\label{tab:param_est}
	\begin{tabular}{p{0.26\textwidth}crrrr}
		\toprule
		\multicolumn{1}{c}{ } & \multicolumn{1}{c}{ } & \multicolumn{2}{c}{SHREVE Model} & \multicolumn{2}{c}{SHRE Model} \\
		\cmidrule(ll{3pt}r{3pt}){3-4} \cmidrule(ll{3pt}r{3pt}){5-6}
		Parameters & Symbols & Mean & 95\% CrI & Mean & 95\% CrI\\
		\midrule
		\multicolumn{6}{c}{Global Parameters} \\
		Intercept & $\mu_0$ & 70.02 & (54.47, 84.21) & 71.22 & (56.83, 84.80) \\
		Slope & $\mu_1$ & -0.30 & (-0.59, 0.02) & -0.30 & (-0.60, 0.04) \\
		Log Residual SD & $\mu_{\phi}$  & 0.35 & (0.05, 0.86) & 0.66 & (0.39, 0.97) \\
		\midrule
		\multicolumn{6}{c}{Subject-Level MGP SD Parameters} \\
		Intercept & $\sigma_{\beta, 11}$ & 16.17 & (15.11, 17.39) & 16.33 & (15.24, 17.57) \\
		Slope & $\sigma_{\beta, 22}$ & 0.94 & (0.87, 1.03) & 1.00 & (0.92, 1.09)  \\
		Log Residual SD & $\sigma_{\beta, 33}$ & 0.45 & (0.42, 0.49) & 0.34 & (0.32, 0.37) \\
		\midrule
		\multicolumn{6}{c}{Subject-Level MGP Lengthscale Parameters} \\
		Intercept & $\ell_{\beta, 1}$ & 5.42 & (4.67, 6.32) & 5.58 & (4.80, 6.51) \\
		Slope & $\ell_{\beta, 2}$  & 4.20 & (3.41, 5.16) & 6.79 & (5.48, 8.46) \\
		Log Residual SDs & $\ell_{\beta, 3}$ & 1.87 & (1.57, 2.24) & 3.71 & (3.02, 4.61) \\
		\midrule
		\multicolumn{6}{c}{Subject-Level MGP Correlation Parameters} \\
		Intercepts/Slopes & $\rho_{\beta, 12}$ & -0.14 & (-0.19, -0.10) & -0.13 & (-0.18, -0.08) \\
		Intercepts/Log Residual SDs & $\rho_{\beta, 13}$ & 0.12 & (0.08, 0.16) & 0.17 & (0.11, 0.22) \\
		Slopes/Log Residual SDs & $\rho_{\beta, 23}$ & -0.21 & (-0.28, -0.14) & -0.24 & (-0.31, -0.17)  \\
		\midrule
		\multicolumn{6}{c}{Visit Effect Parameters} \\
		Lengthscale & $\ell_{v}$ & 3.54 & (3.07, 4.10) &  &  \\
		SD & $\sigma_{v}$ & 1.42 & (1.37, 1.48) &  &  \\
		\bottomrule
	\end{tabular}
\end{table}

Table \ref{tab:param_est} gives posterior means and 95\% credible intervals (CrI) for parameters of interest from the SHREVE and SHRE models. The SHREVE global log residual SD parameter has a smaller posterior mean than SHRE (0.35 vs 0.66 $\mu$m), although CrIs overlap; global intercepts and slopes have similar posterior means and CrIs. The SHREVE subject-level slopes and log residual SDs MGP lengthscales are shorter than for the SHRE model, implying that the spatial correlation of subject-level slopes and log residual SDs decays faster after including visit effects, allowing random effects to vary more across the macula. The SHREVE subject-level MGP SD parameter is larger than from SHRE, meaning the variability of subject-specific residual SDs is higher within a superpixel for the SHREVE model. All other subject-level MGP parameters are similar between the models. Appendix Table \ref{tab:param_est_full} gives posterior means and 95\% CrIs for the population-level MGP parameters. The population-level MGP parameters are similar between the two models.

\begin{figure}[t]
	\centering
	\includegraphics[width=0.95\textwidth]{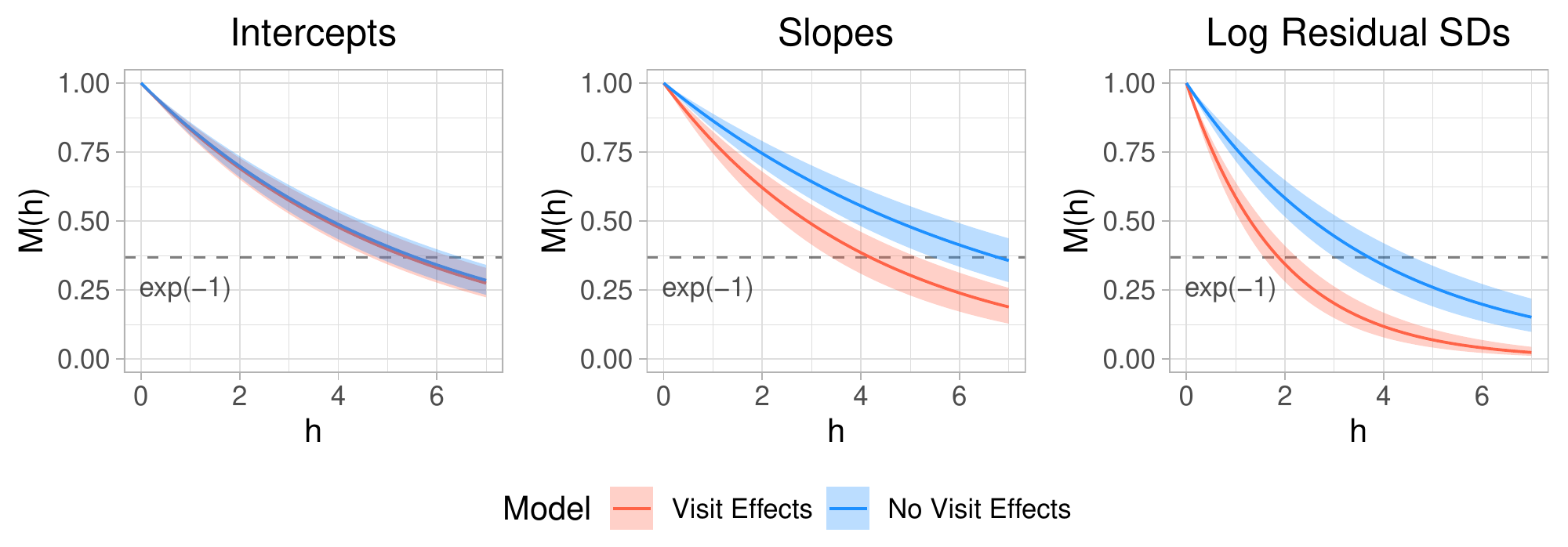}
	\caption{Posterior mean (line) and 95\% pointwise credible intervals (colored bands) of correlation as a function of distance $h$ between superpixels for subject-specific intercepts, slopes, and log residual SDs from the SHREVE (Visit Effects) and SHRE (No Visit Effects) models. The correlations decay faster in the SHREVE model with shorter lengthscales for slopes and log residual SDs. The dashed line indicates where the correlation is $\exp(-1)$ and the distance between superpixels is equal to the lengthscale in the exponential kernel.}
	\label{fig:cor_fxn}
\end{figure}

Figure \ref{fig:cor_fxn} plots spatial correlations $M(h)$ as a function of distance $h$ between superpixels for the SHREVE and SHRE models. At 4.2 units distance, the spatial correlation of subject-specific slopes drops to $\exp(-1) \approx 0.37$ for the SHREVE model but is $\exp(-0.62) \approx 0.54$ for the SHRE model. At 1.9 units distance, the spatial correlation of subject-specific log residual SDs is 0.37 for the SHREVE model but around 0.60 for the SHRE model. The shorter lengthscales in the SHREVE model result in reduced correlation at similar distances between superpixels.

\begin{figure}[t]
\centering
\includegraphics[width=0.75\textwidth]{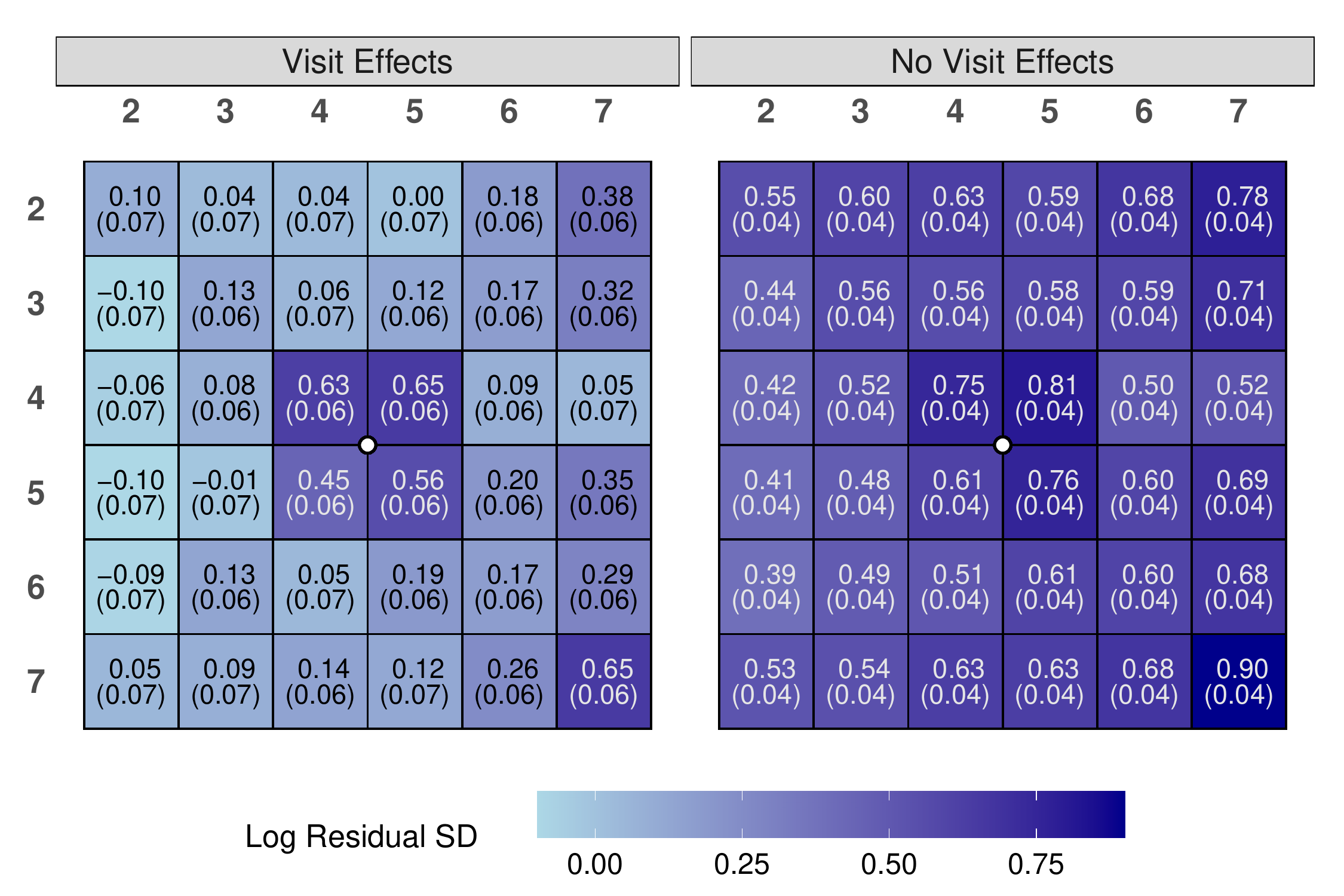}
\caption{Heatmap of the log residual standard deviations (SD) comparing the SHREVE (Visit Effects) and SHRE (No Visit Effects) models. The values shown are the posterior mean (posterior SD) across the 36 superpixels. The log residual SDs from the SHREVE model are uniformly reduced across all superpixels compared to those from the SHRE model. The white dot is the fovea.}
\label{fig:log_sd}
\end{figure}

Figure \ref{fig:log_sd} presents heatmaps of the posterior means and SDs of the log residual SDs from the SHREVE and SHRE models. For most superpixels, the SHREVE model uniformly reduces log residual SDs by approximately 0.5 compared to the SHRE model. The four central superpixels (4.4, 4.5, 5.4, and 5.5) and superpixels in the 7th column have higher log residual SDs and have smaller differences in log residual SDs between the models. SHREVE breaks down measurement error into two components, spatially correlated errors due to the imaging process and general measurement noise. By accounting for visit effects, we reduce residual variance, leading to substantial improvement in model fit.

\begin{figure}[t]
\centering
\includegraphics[width=0.8\textwidth]{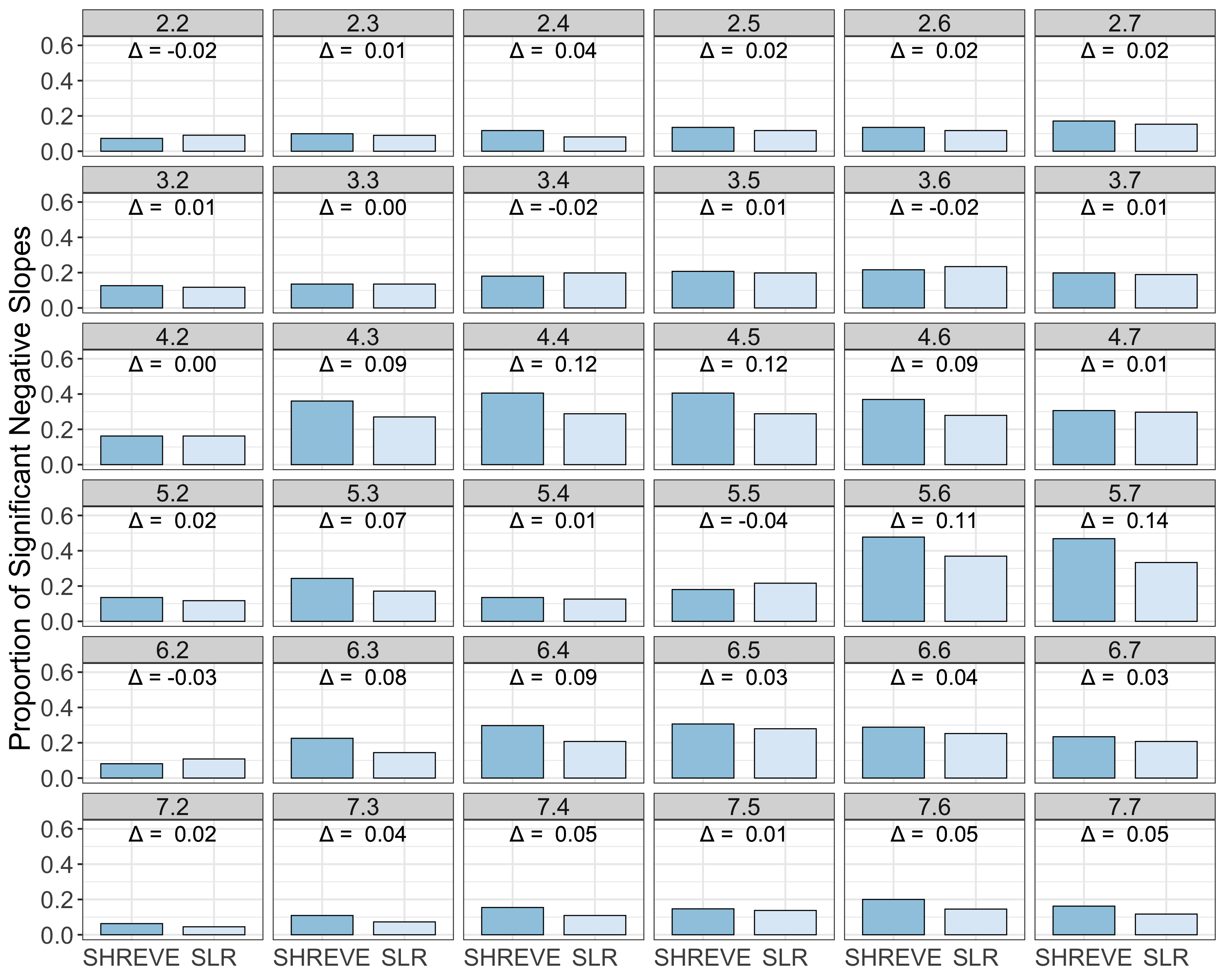}
\caption{Bar charts of the proportion of significant negative slopes detected by the SHREVE model and simple linear regression (SLR) across the 36 superpixels. The difference ($\Delta = \mbox{SHREVE} - \mbox{SLR}$) in proportion is labeled at the top of each subplot. Across all locations, the SHREVE model detects a higher proportion of significant negative slopes (21.4\% vs 18.0\%) than SLR.}
\label{fig:neg_prop}
\end{figure}

We compare subject-specific slopes estimated from the SHREVE model to those estimated using SLR. We declare a slope to be significantly negative or positive when the upper bound or lower bound of the 95\% CrI is less than or greater than 0, respectively. Across the 3,990 subject-superpixel profiles, the SHREVE model detects a higher proportion of significant negative slopes (21.4\% vs 18.0\%) and lower proportion of significant positive slopes (3.1\% vs 4.3\%) as compared to SLR. Figure \ref{fig:neg_prop} shows the proportion of significant negative slopes by superpixel, and Appendix Figure \ref{fig:pos_prop} shows the proportion of significant positive slopes by superpixel. The SHREVE model detects 10\% more significant negative slopes in 6 of 36 superpixels and 5\% less significant positive slopes in 5 of 36 superpixels. Because glaucoma is an irreversible disease, GCC thicknesses are not expected to increase over time. These findings indicate SHREVE is more sensitive in detecting worsening slopes and possibly reduces false positive rates as compared to SLR.

\section{Discussion} \label{discussion}

We motivate and develop a Bayesian hierarchical model with population- and subject-level spatially varying coefficients and show that including visit effects reduces error in predicting future observations and greatly improves model fit. In current practice, ophthalmologists use SLR to assess slopes for individual subject-superpixel profiles, using information from only a single subject and location at a time. To better estimate subject-specific slopes, we include information from the whole cohort; explicitly model the correlations between subject-specific intercepts, slopes, and log residual SDs; allow population parameters and random effects to be spatially correlated; and account for visit-specific spatially correlated errors. Using information from the entire cohort, our proposed model leads to decreased noise in estimating subject-specific slopes, having smaller posterior SDs in 79\% of subject-superpixel slopes as compared to SLR.

There are many sources of error in obtaining the GCC thickness measurements from OCT scans. By separating measurement errors into visit-specific spatially correlated errors and other measurement noise, we are better able to detect eye-superpixels where GCC thicknesses are progressing most rapidly. Our approach will help identify progression of glaucoma for more individualized treatment plans.

Other methods for modeling spatial variation over discrete locations include CAR models, where random effect distributions are conditional on some neighboring values \citep{betz2013spatial,berchuck2019diagnosing}. Instead, we model spatial correlation between all locations with GPs, where the spatial correlation depends only on the distance between any two locations. In addition to our a priori specification of $\nu = \tfrac{1}{2}$, we fit our model using Mat\'ern correlation functions with $\nu = \tfrac{3}{2}$, $\nu = \tfrac{5}{2}$, and $\nu = \infty$ (squared exponential kernel, \citealt{williams2006gaussian}). These early exploratory analyses had difficulty in MCMC convergence. One limitation of using GPs is the increasing difficulty in fitting when the number of locations is large. Fitting GP models involves matrix inversion which increases computational complexity in cubic order with the number of locations. When the number of locations is too large, approximations for the processes could be considered \citep{banerjee2008gaussian}. Nonetheless, we expect these model developments will benefit ophthalmologists as they seek to better estimate subject-specific slopes from structural thickness measurements.

We developed the current model specifically for GCC macular thickness measurements. Of further interest is to simultaneously model all the inner retinal layers that make up GCC to identify which sublayers may be worsening faster than others while accounting for between-layer correlations. Future extensions of the SHREVE model could include working with multivariate outcomes, which may pose additional computational challenges.

%
\begin{appendix}
	
\beginsupplementA
	
\section{Convergence assessment of the SHREVE model}
\begin{figure}[tp!]
	\centering
	\includegraphics[width=0.8\textwidth]{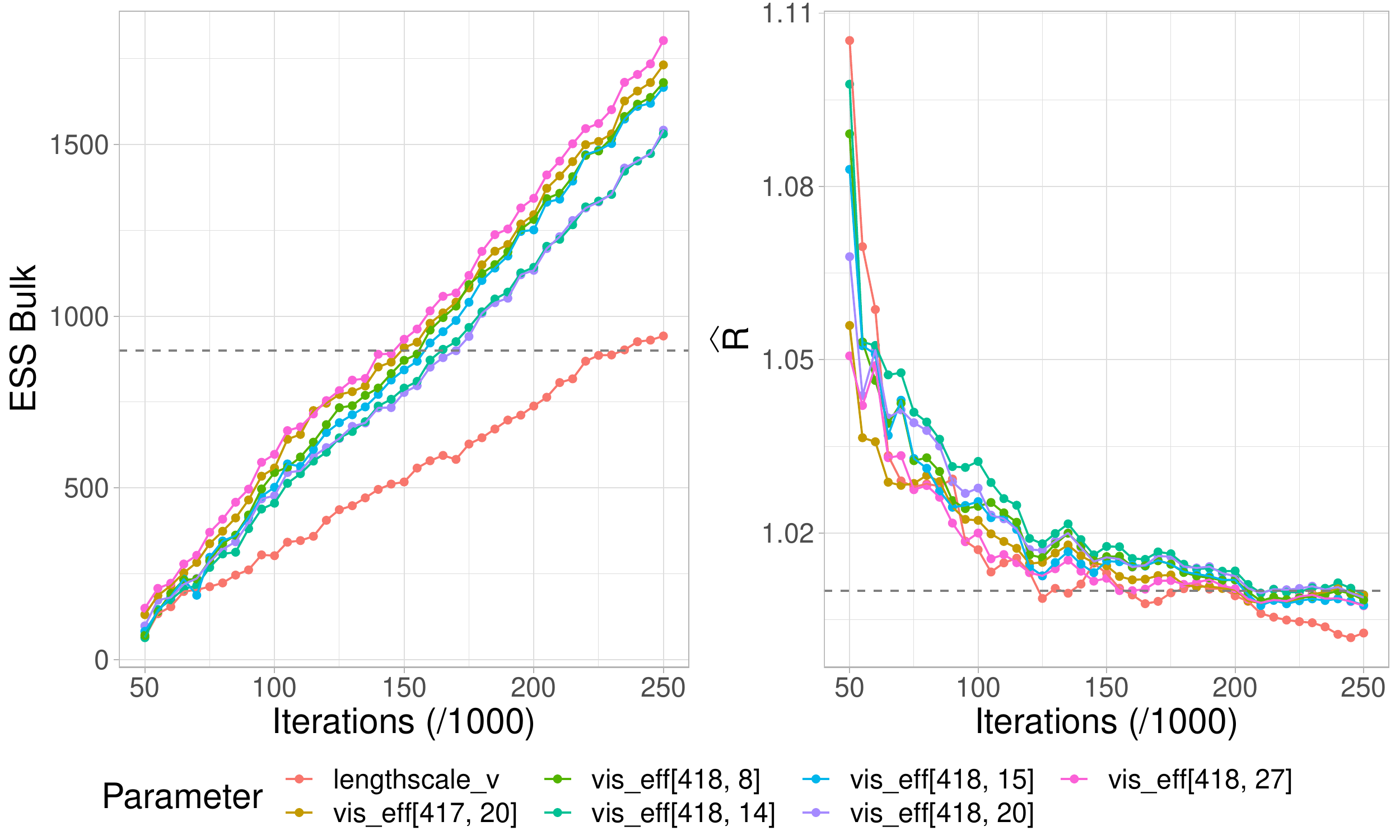}
	\caption{Plots of efficiency per iteration of the bulk effective sample size (ESS) on the left and potential scale reduction factor $\widehat{R}$ on the right for the 7 parameters with the largest $\widehat{R}$ from the SHREVE model. The bulk ESS increases linearly with increasing iterations while $\widehat{R}$ decreases exponentially with increasing iterations. The bulk ESSs were all greater than 100 per chain and $\widehat{R}$ were all less than 1.01.}
	\label{fig:ess}
\end{figure}

We provide more details on convergence of the SHREVE model as mentioned in Section 3.5. Following \citet{vehtari2021rank}'s recommendation on assessing convergence, we monitored the potential scale reduction factor $\widehat{R}$ and the bulk and tail effective sample sizes (ESS) for all model parameters. We found $\widehat{R}$ was less than 1.01 and bulk and tail ESS were all greater than 100 per chain for all parameters, leading us to believe that our MCMC has converged satisfactorily. Figure \ref{fig:ess} shows the efficiency per iteration of the bulk ESS and potential reduction factor $\widehat{R}$ of 7 model parameters with the largest $\widehat{R}$ in the SHREVE model. The bulk ESS increases linearly with increasing iterations indicating that the relative efficiency is constant over different numbers of draws. $\widehat{R}$ decreases exponentially with increasing iterations and are all less than 1.01. Table \ref{tab:convergence} gives the sampling efficiency of all model parameters in terms of bulk and tail ESS and $\widehat{R}$.

\begin{table}[tp!]
	\centering
	\tabularnewline
	\caption{The mean minimum/maximum bulk effective sample size (ESS), tail ESS, and potential scale reduction factor $\widehat{R}$ for the SHREVE model broken down by parameter types. The first column gives the parameter types: Hyperparameters include global parameters and MGP hyperparameters; Population-level includes the population-level superpixel intercepts, slopes, and log residual SDs; Intercepts, Slopes, and Log Residual SDs include the subject-level intercepts, slopes, and log residual SDs across all locations, respectively; and Visit Effects include the visit effects across all locations. The second column gives the number of parameters summarized.}
	\label{tab:convergence}
	\begin{tabular}{lrrrrccl}
		\toprule
		\multicolumn{2}{c}{ } & \multicolumn{2}{c}{Bulk ESS} & \multicolumn{2}{c}{Tail ESS} & \multicolumn{2}{c}{$\widehat{R}$} \\
		\cmidrule(l{3pt}r{3pt}){3-4} \cmidrule(l{3pt}r{3pt}){5-6} 
		\cmidrule(l{3pt}r{3pt}){7-8}Parameter & \# & Mean & Min & Mean & Min & Mean & Max\\
		\midrule
		Hyperparameters & 23 & 12545.8 & 942.3 & 14152.8 & 1951.3 & 1.001 & 1.003\\
		Population-level & 108 & 11519.7 & 3706.8 & 15610.8 & 7478.1 & 1.001 & 1.002\\
		Intercepts & 3990 & 4406.5 & 1663.8 & 9644.9 & 3585.9 & 1.002 & 1.007\\
		Slopes & 3990 & 3757.7 & 1602.4 & 8559.5 & 3597.6 & 1.002 & 1.007\\
		Log Residual SDs & 3990 & 16268.4 & 9311.0 & 18271.6 & 14123.2 & 1.000 & 1.002\\
		Visit Effects & 29179 & 4963.8 & 1505.0 & 10572.5 & 3475.9 & 1.002 & 1.009\\
		\bottomrule
	\end{tabular}
\end{table}

\section{Additional results for AGPS analysis}

\beginsupplementB

\begin{figure}[htp!]
	\centering
	\includegraphics[width=0.8\textwidth]{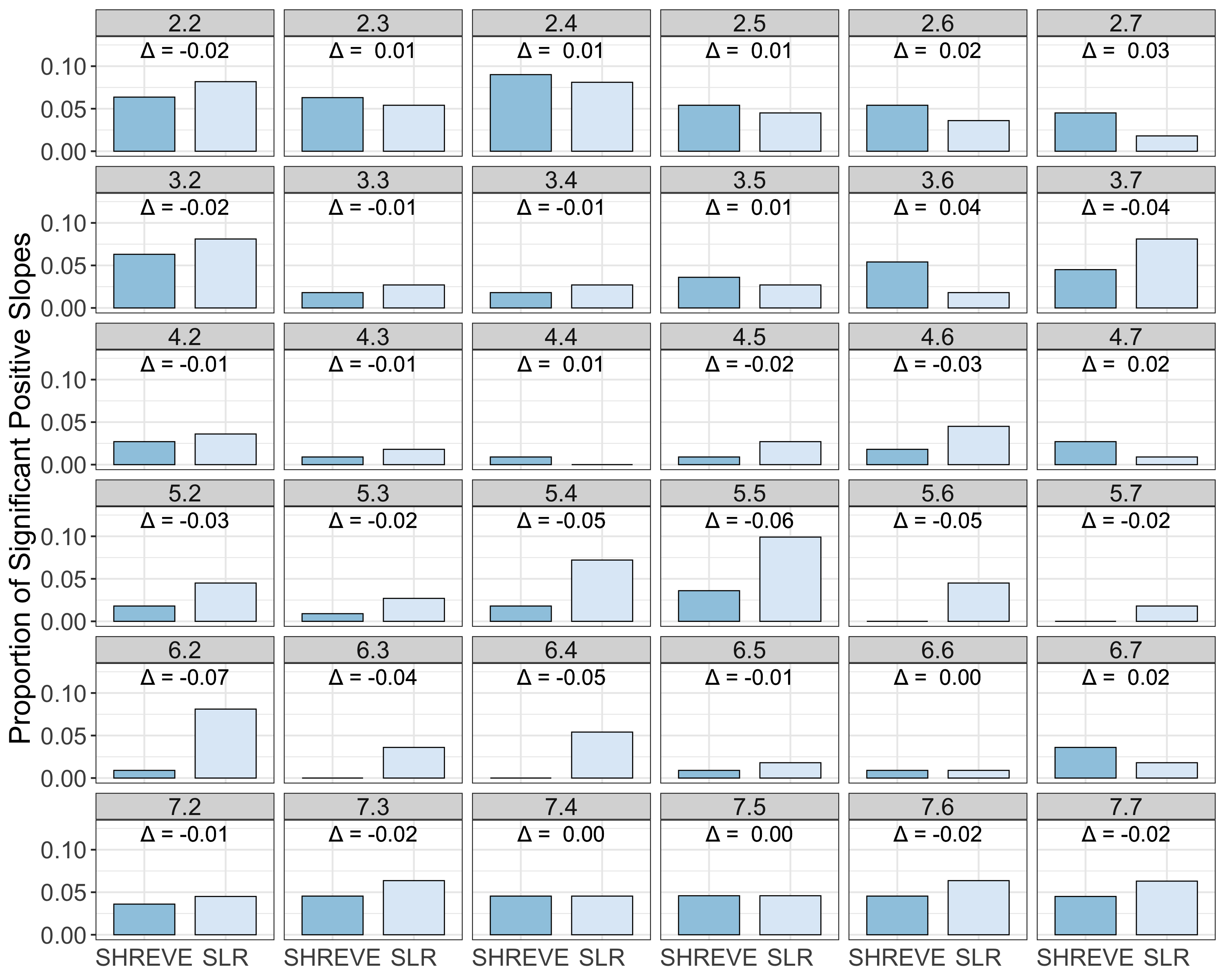}
	\caption{Bar charts of the proportion of significant positive slopes detected by the SHREVE model and simple linear regression (SLR) across the 36 superpixels. The difference ($\Delta = \mbox{SHREVE} - \mbox{SLR}$) in proportion is labeled at the top of each subplot. Across all locations, the SHREVE model detects a lower proportion of significant positive slopes (3.1\% vs 4.3\%) than SLR.}
	\label{fig:pos_prop}
\end{figure}

We provide additional results mentioned in Section 4. Table \ref{tab:param_est_full} presents the posterior mean and 95\% CrIs for population-level MGP parameters for the SHREVE and SHRE models. All population-level MGP parameter posterior means and 95\% CrIs are similar between the models. Figure \ref{fig:pos_prop} plots the proportions of significant positive slopes for the SHREVE model and SLR in each of the 36 superpixels. Across all locations, the SHREVE model detects a lower proportion of significant positive slopes (3.1\% vs 4.3\%) than SLR.

\begin{table}[htp!]
	\centering
	\tabularnewline
	\caption{Posterior mean and 95\% credible interval (CrI) for population-level multivariate Gaussian process parameters for the SHREVE and SHRE models.}
	\label{tab:param_est_full}
	\begin{tabular}{p{0.31\textwidth}crrrr}
		\toprule
		\multicolumn{1}{c}{ } & \multicolumn{1}{c}{ } & \multicolumn{2}{c}{SHREVE Model} & \multicolumn{2}{c}{SHRE Model} \\
		\cmidrule(ll{3pt}r{3pt}){3-4} \cmidrule(ll{3pt}r{3pt}){5-6}
		Parameters & Symbols & Mean & 95\% CrI & Mean & 95\% CrI\\
		\midrule
		\multicolumn{6}{c}{Population-Level MGP SD Parameters} \\
		Intercept & $\sigma_{\alpha, 11}$ & 13.68 & (9.27, 21.13) & 13.31 & (9.06, 20.55) \\
		Slope & $\sigma_{\alpha, 22}$ & 0.31 & (0.20, 0.54) & 0.32 & (0.21, 0.56) \\
		Log Residual SD & $\sigma_{\alpha, 33}$ & 0.36 & (0.19, 0.79) & 0.22 & (0.11, 0.47) \\
		\midrule
		\multicolumn{6}{c}{Population-Level MGP Lengthscale Parameters} \\
		Intercept & $\ell_{\alpha, 1}$ & 3.56 & (1.32, 8.83) & 3.27 & (1.20, 8.16) \\
		Slope & $\ell_{\alpha, 2}$ & 2.66 & (0.88, 8.33) & 2.82 & (0.94, 8.90) \\
		Log Residual SD & $\ell_{\alpha, 3}$ & 4.68 & (0.75, 19.68) & 6.43 & (0.98, 26.16) \\
		\midrule
		\multicolumn{6}{c}{Population-Level MGP Correlation Parameters} \\
		Intercepts/Slopes & $\rho_{\alpha, 12}$ & -0.42 & (-0.68, -0.13) & -0.42 & (-0.67, -0.12) \\
		Intercepts/Log Residual SDs & $\rho_{\alpha, 13}$ & -0.30 & (-0.57, -0.02) & -0.28 & (-0.55, 0.01) \\
		Slopes/Log Residual SDs & $\rho_{\alpha, 23}$ & -0.11 & (-0.42, 0.20) & -0.06 & (-0.37, 0.24) \\
		\bottomrule
	\end{tabular}
\end{table}

\end{appendix}

\begin{funding}
	This work was supported by an NIH R01 grant (R01-EY029792), an unrestricted Departmental Grant from Research to Prevent Blindness, and an unrestricted grant from Heidelberg Engineering. AJH was supported by NIH grant K25 AI153816, NSF grant DMS 2152774, and a generous gift from the Karen Toffler Charitable Trust.
	
\end{funding}

\bibliographystyle{imsart-nameyear} 
\bibliography{ref_list}       


\end{document}